\documentclass[12pt,oneside,final]{thesis}

\usepackage[english]{babel}
\usepackage{blindtext}

\usepackage{cite}
\usepackage{amsmath,amsfonts}

\usepackage[final]{pdfpages}
\usepackage{amsthm}

\theoremstyle{definition}

\theoremstyle{remark}

\theoremstyle{remark}

\usepackage{etoolbox}
\usepackage{framed}
\newcommand\enclosebox[2]{%
  \BeforeBeginEnvironment{#1}{\begin{#2}}%
  \AfterEndEnvironment{#1}{\end{#2}}%
}
\enclosebox{definition}{leftbar} 
\enclosebox{remark}{leftbar} 

\usepackage{mypackage}
\usepackage{graphicx}
\usepackage{fixltx2e}
\usepackage{array}
\usepackage{wrapfig} 
\usepackage{newcent}

\usepackage[linktocpage=true]{hyperref} 

\usepackage{upgreek} 
\usepackage{bm}
\usepackage{hyperref}
\usepackage{setspace}
\usepackage{booktabs}
\usepackage{multirow}
\usepackage{longtable}
\usepackage[font=singlespacing, labelfont=bf]{caption}

\usepackage{enumitem}
\newlist{inlinelist}{enumerate*}{1}
\setlist*[inlinelist,1]{%
  label=(\arabic*),
}

\usepackage{fancyhdr}    

\fancypagestyle{plain}{
\fancyhf{} 
\fancyfoot[C]{\thepage} 

}

\usepackage{amsthm}

\begin{document}

\title{{\bf\Large{ Cosmological Perturbations in a Universe with a Domain Wall Era }}}
\author{{\bf Cyrus Faroughy}}
\degreemonth{December}
\degreeyear{2017} 
\dissertation
\doctorphilosophy
\copyrightnotice


\begin{frontmatter}

\maketitle

\begin{abstract}
Topologically protected sheet-like surfaces, called \textit{domain walls}, form when the potential of a field has a discrete symmetry that is spontaneously broken. Since this condition is commonplace in field theory, it is plausible that many of these walls were produced at some point in the early universe. Moreover, for potentials with a rich enough structure, the walls can join and form a (at large scales) homogeneous and isotropic network that dominates the energy density of the universe for some time before decaying.
\\
\indent In this thesis, we study the faith of large scale perturbations in a cosmology with a short period of domain wall dominance. We start with some background material introducing the unperturbed universe, the physics of inflation and the basics on domain walls in cosmology. We also review the theory of cosmological perturbations in a medium with non-zero anisotropic stress. We then move on to our main result: treating the domain wall network as a relativistic elastic solid at large scales, we show that the perturbations that exited the horizon during inflation get \textit{suppressed} during the domain wall era, before re-entering the horizon. This power suppression occurs because, unlike a fluid-like universe, a solid-like universe can support sizable anisotropic stress gradients across large scales which effectively act as mass for the scalar and tensor modes. Interestingly, the amplitude of the primordial scalar power spectrum can be closer to one in this cosmology and still give the observed value of $10^{-9}$ today. As a result, the usual bounds on the energy scale of inflation get relaxed to values closer to the (more natural) Planck scale.  
\\
\indent In the last part of this thesis, as an existence proof, we present a hybrid inflation model with $N$ ``waterfall'' fields that can realize the proposed cosmology. In this model, a domain wall network forms when an approximate $O(N)$ symmetry gets spontaneously broken at the end of inflation, and for $N\geq 5$, we show that there is a region in parameter space where the network dominates the energy density for a few e-folds before decaying and reheating the universe.
\\
\vfill
\noindent {\bf{Primary Reader and Advisor:}} David. E. Kaplan\\
{\bf{Secondary Readers:}} Ibou Bah, Petar Maksimovic, Tobias Marriage, Yannick Sire
\end{abstract}

\begin{acknowledgment}

I would like to thank David Kaplan for all his guidance and for helping me better understand many of the concepts discussed in this thesis. I also want to thank everyone in the particle theory group at the Johns Hopkins University, in particular, my office-mates Hongbin Chen and Charles Hussong, for all the stimulating discussions.
\\
\indent Finally, I am very grateful to my parents and brother for their love and support, and Bonnie for her constant encouragement and for helping me have a well balanced life in Baltimore.

\end{acknowledgment}

\begin{dedication}
 
To my physicist father who -- since day one -- taught me the most valuable lesson: to be curious about everything in Nature, from the little ants on the ground to the big moon in the sky. I would not be where I am today without his constant support throughout every step of my career.

\end{dedication}

\tableofcontents

\listoftables

\listoffigures

\end{frontmatter}

\chapter{Introduction}
\label{ch:intro}
\chaptermark{Introduction}

\makeatletter
\let\@currsize\normalsize
\makeatother

\begin{figure}[t]
\center{
\includegraphics[scale=0.6]{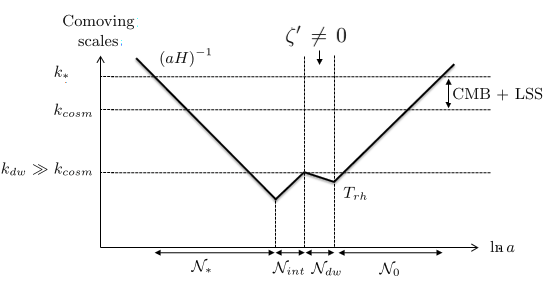}
\caption{Evolution of the comoving Hubble radius, $\mathcal{H}^{-1}= (a H)^{-1}$, where $a$ is the scale factor and $H$ is the Hubble parameter, in a cosmology with a domain wall dominated era. Here, $\N_{*}$ is the number of e-folds from horizon-exit of the pivot scale $k_*\simeq0.05$ Mpc$^{-1}$ to the end of inflation ($w=-1$), $\N_{int}$ is the number of e-folds elapsed during a radiation-dominated phase after inflation but before domain walls dominate  ($w=1/3$), $\N_{dw}$ is the number of e-folds during the domain wall era ($w=-2/3$), and $\N_0$ is the number of e-folds from reheating until the $k_*$ perturbations would re-enter the horizon. $T_{rh}$ is the temperature at reheating. $k_{cosm}\sim1$ Mpc$^{-1}$ is the mode corresponding to the smallest cosmological scale currently probed by experiments, and $k_{dw}\gg k_{cosm}$ is the mode exiting the horizon when the domain walls start to dominate. Lastly, $\zeta$ is the curvature perturbation which, as we will show in this thesis, decays at superhorizon scales during domain wall dominance.\label{fig:CosmicHistory}}}
\end{figure}

The cosmic inflation paradigm \cite{Guth:1980zm, Linde:1981mu, Starobinsky:1980te} is a theory which postulates that a period of exponential expansion of space existed right before the universe became dense and hot. Remarkably, this rapid expansion period in the very early universe is enough to explain why we observe it to be so flat and homogeneous at large scales today, and at the same time, it explains the origin of the perturbations needed to seed the formation of large-scale structure, like galaxies and galaxy clusters. When combined with the standard big bang model, which predicts how the hot and dense (radiation-dominated) universe expands and cools until today, one can explain almost everything we observe in the universe (see \cite{KolbTurner} for a review).
\\
\indent In its simplest incarnation -- single-field slow-roll inflation -- the exponential expansion is driven by a scalar field that permeates all of space, called the \textit{inflaton}, whose energy density is mostly stored as potential energy; the kinetic contribution is always small and the field is said to ``slow-roll'' down its potential. Then, because of the rapid expansion of space, the quantum fluctuations in the inflaton field get amplified to cosmological scales and eventually leave the causal horizon (we review this in Section \ref{sec:InfCosmo}). When inflation ends (and in the canonical scenario, the energy density quickly becomes radiation dominated), the expansion of the universe slows down dramatically and these \textit{cosmological perturbations} re-enter the horizon; seeding the gravitational instabilities necessary for structure formation. 
\\
\indent In the standard lore, the perturbations generated during inflation remain ``frozen'' outside the horizon (they are constant). This means that one only needs to calculate the amplitude of these perturbations at the time of horizon exit to make predictions today. However -- and this is one of the main points of this thesis -- for this to be always true one must assume that,  while the perturbations are outside the horizon, every dominant component in the post-inflationary universe behaves like a \emph{perfect fluid} at large scales. This assumption is often overlooked since in the standard cosmological model, shortly after inflation ends, the universe first becomes dominated by radiation, then matter, then vacuum energy; all of which have vanishing anisotropic stress gradients at large scales and can therefore be described as perfect fluids. Moreover, though extensions to this model exist in the literature (see e.g., \cite{Munoz:2014eqa}), they typically involve adding an intermediate epoch (parameterized by some effective equation state $w>-1/3$) that is also assumed to behave like a perfect fluid at large scales. Since we know little about the universe between the end of inflation and the onset of the radiation era (see \cite{Abbott:1982hn, Kofman:1994rk, Kofman:1997yn, Greene:1997ge, Amin:2014eta}, for different examples of how this ``reheating'' transition can occur), it is worth exploring more general cases which go beyond the perfect fluid approximation.
\\
\indent In this thesis we study a cosmology where a \textit{domain wall network} -- which behaves more like a \textit{relativistic elastic solid} at large scales -- dominates the energy density for a short period in the early universe, not long after inflation ends (as we will mention shortly, this period must be short and must occur early in order to avoid current observational bounds). At the microscopic level, this happens if the potential of a field has a discrete symmetry that is spontaneously broken: domains walls, which are topologically protected sheet-like surfaces, form and divide the universe into domains populated at random by one of the available vacua \cite{Kibble:1976sj, Zeldovich}. Since this condition is commonplace in field theory, it is quite possible that many of these walls were produced in the early universe. Moreover, if the vacuum manifold of the theory is rich enough, the domain walls can join together into a metastable, (and at large scales) homogeneous and isotropic network that can dominate the energy density for some number of e-folds before decaying and reheating the universe. As we will see in Section \ref{sec:Review}, the equation of state during domain wall dominance (assuming frustration) is $w=-2/3$ \cite{KolbTurner, CosmicStrings}, so the universe enters a second period of accelerated expansion. The evolution of the comoving Hubble radius in this cosmology, is sketched in Fig.(\ref{fig:CosmicHistory}).
\\
\indent The study of topological defects in cosmology, such as strings and walls, is not new \cite{Kibble:1976sj, Zeldovich}. In fact, in the 1990's, a lot of work was done to show that defects could be primary sources of anisotropy in the Cosmic Microwave Background \cite{CosmicStrings}, therefore providing an alternative to the inflation paradigm. However, these alternatives have since been shown to be incompatible with data: unless the field constituents are light ($\sim$ MeV), long-lived domain walls that formed early are ruled out and therefore need to decay \cite{Zeldovich, Lazanu:2015fua}.
Domain wall networks produced at late-times in the universe have also been considered; since the equation of state for a network lies in the range $-2/3<w<-1/3$, it could potentially contribute to dark energy. However, this possibility was also ruled out from the commonly quoted upper bound $w < -0.78$ at $95\%$ C.L (though this bound has been contested in \cite{Conversi:2004pi}). In the cosmology we propose, because the network has a characteristic length scale that is always much smaller than the horizon, and because it only lives for a few e-folds shortly after inflation, none of these constraints apply.
\\
\indent Let us summarize the main results of this thesis. Treating the domain wall network as a relativistic elastic solid at large scales, we show in Chapter \ref{ch:DWCosmo} that cosmological perturbations \textit{decay} outside the horizon during domain wall dominance. This means that for the cosmology shown in Fig.(\ref{fig:CosmicHistory}) (or any cosmology with a domain wall era), the amplitude of the perturbations generated during inflation now has to be larger in order to match the amplitude observed today. More specifically, after decomposing the cosmological perturbations into their scalar ($\zeta$) and tensor ($h$) components (as done in Chapter \ref{ch:CosmoPert}), we find that the ratio between the power spectra at large scales \textit{today}, which we denote by $\mathcal{P}_{\zeta,0}$ and $\mathcal{P}_{h,0}$, and their \textit{primordial} values at horizon exit, which we denote by $\mathcal{P}_{\zeta}$ and $\mathcal{P}_{h}$, is exponentially suppressed as
\begin{equation}\label{eq:AlteredPower}
\frac{\mathcal{P}_{\zeta,0}}{\mathcal{P}_{\zeta}}= \frac{\mathcal{P}_{h,0}}{\mathcal{P}_{h}} \propto e^{-\frac{5}{2} \N_{dw}}
\end{equation}
where $\N_{dw}$ is the number of e-folds elapsed during domain wall dominance. In other words, the primordial amplitude of the scalar and tensor power spectra is a factor of $e^{\frac{5}{2}\N_{dw}}$ larger than the amplitude today, which for scalars is $\mathcal{P}_{\zeta,0}(k_*)\sim 10^{-9}$ (as measured by the Planck collaboration \cite{Planck} at the pivot scale $k_*\simeq0.05$ Mpc$^{-1}$). Eq.(\ref{eq:AlteredPower}) is the main result of this thesis.
\\
\indent Even though the tensor-to-scalar ratio $r_*\equiv\mathcal{P}_{h}(k_*)/\mathcal{P}_{\zeta}(k_*)$ does not change in the cosmology with the domain walls, the observational bound \cite{Planck} $r_*< 0.07$ combined with Eq.(\ref{eq:AlteredPower}), places the following bound on the Hubble scale in single-field slow-roll inflation models:
\begin{equation}\label{eq:energyscalebound}
H_i \lesssim 10^{14}\times e^{\frac{5}{4} \N_{dw}} \text{GeV} 
\end{equation} 
We conclude that the main effect of the domain wall network is to exponentially suppress the power spectra and that, interestingly, this suppression relaxes the current bounds on $H_i$ by a factor of $e^{\frac{5}{4} \N_{dw}}$. The only bound on $H_i$ would come from considerations of eternal inflation.  Saturating this bound, the inflationary scale can take values close to the Planck scale. As a result, the parameter space of single-field inflation models is shifted relative to the standard cosmological model. It is worth noting that the evolution of perturbations at superhorizon scales does not represent a violation of causality since the effect that causes it is entirely local -- it is due to the rigid nature of solids which, as we will see, can support sizable anisotropic stress gradients across large scales (see \cite{Sigurdson}, \cite{Junpu}) effectively acting as mass for the perturbations.
\\
\indent The main result, Eq.(\ref{eq:AlteredPower}), is independent of the model used to produce the domain wall network, as long as the network at large scales remains homogeneous and isotropic throughout its lifetime. Nevertheless, we present in Chapter \ref{ch:Model}, as an existence proof, a hybrid inflation model with $N$ waterfall fields where a domain wall network forms at the end of inflation, when an approximate $O(N)$ symmetry gets spontaneously broken. For $N\geq5$, we show that the network can dominate the energy density for a few e-folds before decaying. When the walls decay, the universe reheats and the standard Big Bang cosmology follows. This model can therefore realize the cosmology depicted in Fig.(\ref{fig:CosmicHistory}).
\\
\indent The structure of this thesis is as follows. In Chapter \ref{ch:Background}, we review the basics of cosmology, inflation and domain wall networks. In Chapter \ref{ch:CosmoPert}, we review the linear theory of cosmological perturbations in a universe with a homogeneous and isotropic background and which supports anisotropic stress gradients at large scales. In Chapter \ref{ch:ElasticBodies}, we review the linearized theory of perturbations in Newtonian and relativistic elastic solids. The rest of the thesis contains the bulk of our contributions to the literature. In Chapter \ref{ch:EvoPert}, after obtaining the superhorizon evolution equations of cosmological perturbations in a solid-like universe, we derive the matching conditions that need to be imposed on perturbations transitioning between fluid-like and solid-like phases. In Chapter \ref{ch:DWCosmo}, we use the results from the previous chapter to obtain expressions for the late-time scalar and tensor power spectra in the cosmology with a domain wall era. We briefly discuss non-Gaussianities. In Chapter \ref{ch:Model}, we discuss the origin and collapse of the domain wall network in the context of a field theory model which can realize the proposed cosmology. We conclude in Chapter \ref{ch:Conclusion}.
\chapter{Background Material}
\label{ch:Background}
\chaptermark{Background Material}
In this section, after establishing part of the notation and conventions used in this thesis, we review the basics of (i) the unperturbed universe (i) the physics of inflation (iii) domain wall networks in cosmology. This will serve as a basis for the rest of the thesis, but can be skipped if the reader is already familiar with these topics.

\section{Notation}
Throughout this thesis, we always set $ c = \hbar = 1$. By ``Planck mass'' we always mean the \textit{reduced Planck mass}
$M_p = (8\pi G)^{-1/2}\simeq 2.4\times 10^{18}$GeV, where $G$ is Newton's constant.
\\
\indent We chose to work with the $(+,-,-,-)$ metric signature. Greek indices take the values $\mu, \nu = 0,1,2,3$ whereas Latin indices take the values $i,j=1,2,3$. We use the letter $\tau$ to denote \textit{conformal time} and $t$ to denote physical time. Unless otherwise specified, primes represent derivatives with respect to $\tau$ and overdots represent derivatives with respect to $t$. 
\\
\indent We use $\mathcal{P}_X(k)$ to refer to the \textit{dimensionless power spectrum} of some perturbation $X$. In Fourier space, it is given by
$$\mathcal{P}_X(k) = \frac{k^3}{2\pi^2} P_X(k); \qquad\qquad \left\langle X_{\vec{k}} X_{\vec{k}^{\prime}}\right\rangle = (2\pi)^3\delta(\vec{k} + \vec{k}^{\prime}) P_X(k)$$
Lastly, when discussing cosmological perturbation theory (except for quantities which use established symbols to denote their perturbations) we use $\bar{Y}$ to denote the background and $\delta Y$ to denote the perturbation of the full quantity $Y = \bar{Y} + \delta Y$.  

\section{The Unperturbed Universe}\label{sec:Unperturbed}
At large scales, our universe is homogeneous and isotropic. It is also spatially flat. Therefore, at the background level (i.e., after averaging over spatial fluctuations), the expanding universe is described by the Friedmann-Robertson-Walker (FRW) metric:
\begin{equation}\label{eq:FRWmetric}
ds^2 = dt^2 - a(t)^2(dx^2 + dy^2 + dz^2)
\end{equation}
where $a(t)$ is the scale factor parameterizing the expansion of the universe and $t$ is the cosmological time (i.e. the proper time measured by a free-falling observer). The rate of expansion is given by the Hubble parameter $H \equiv \dot{a}/a$.
\\
\indent The matter sector is described by a relativistic \emph{perfect fluid}. Before recombination, for example, the Thomson scattering processes between photons and electrons, and the Coulomb interactions between the electrons and baryons, were very efficient. Thus, we can safely assume that the Thomson mean free path $\tau_m$ during that period was negligible compared to the Hubble scale\footnote{Corrections of order $\tau_m/H^{-1}$ and higher are very small and thus neglected.} $H^{-1}$. The perfect fluid is completely characterized by its energy density $\rho$ and isotropic pressure $P$. Indeed, if the mean free path was comparable to the size of the horizon, then it would be possible for an observer to receive different fluxes with different velocities from different directions and, as a result, there would be a net pressure along a particular direction. This would lead to \emph{anisotropic stress} and the fluid would no longer be perfect\footnote{This occurs, for example, after neutrino decoupling.}.
\\
\indent Let us consider the energy-momentum tensor $T^{\mu\nu}$. In general one can interpret this tensor as the flux of the $\mu$-th component of 4-momentum across a surface of constant $x^\nu$. In terms of components, we have
\begin{itemize}
\item $T^{00}$: Energy density (energy flux across constant time surface).
\item $T^{0i}$: Heat conduction (energy flux across constant $x^i$ surface).
\item $T^{i0}$: Momentum density ($i$-th momentum across constant time surface).
\item $T^{ij}$: Stress ($i$-th momentum across a constant $x^j$ surface).
\end{itemize}
A perfect fluid has by definition (1) no heat conduction (2) no viscosity, so $T^{0i}=T^{i0}=0$, and (3) $T^{ij}(\vec{x} =0)\propto \delta^{ij} \propto g_{ij}(\vec{x} =0$), where $g_{ij}$ denotes the metric. This follows from isotropy: an observer at rest relative to any point within the fluid would see the same thing in all different directions. Homogeneity then requires that all proportionality coefficients be functions of time only. We thus have
$$T_{00} = \rho(t), \quad\quad T_{i0}=0, \quad\quad T_{ij}= -P(t) g_{ij}(t,\vec{x})$$
For an observer comoving with the perfect fluid, the stress energy tensor with mixed indices takes the following nice form
\begin{equation}\label{eq:Tmunumatrix}
T^\mu_\nu = 
\begin{bmatrix}
    \rho & 0 & 0 & 0 \\
    0 & -P & 0 & 0  \\
    0 & 0 & -P & 0 \\
    0 & 0 & 0 & -P
\end{bmatrix}
\end{equation}
One can then boost to any frame, moving at speed $\mathbf{v}$ relative to the rest frame, using the Lorentz transformation matrix. Defining $u^\mu=\gamma(1,\mathbf{v})$ such that $u^\mu u_\mu=1$ and $\gamma=1/\sqrt{1-\mathbf{v}^2}$, the usual covariant form then follows
\begin{equation}
T^\mu_\nu=(\rho + P)u^\mu u_\nu - P \delta^\mu_\nu
\end{equation}
Physically, $u^\mu=dx^\mu/ds$ is the relative four-velocity between the fluid and the observer, and $\rho$ and
$P$ are the energy density and pressure in the rest-frame of the fluid, respectively. Note that for $u^\mu = (1,0,0,0)$ (the four-velocity of a comoving observer) we recover the result in Eq.(\ref{eq:Tmunumatrix}).
\\
\indent The evolution of the energy density and pressure with respect to time can be found from the conservation of energy and momentum condition
$$\nabla_\mu T^\mu_\nu = \partial_\mu T^\mu_\nu + \Gamma^\mu_{\mu \lambda} T^\lambda_\nu - \Gamma^\lambda_{\mu \nu} T^\mu_\lambda=0$$
where $\Gamma^{\mu}_{\alpha\beta}\equiv\frac{1}{2}g^{\mu\lambda}\left(\partial_\alpha g_{\beta\lambda} + \partial_\beta g_{\alpha\lambda} - \partial_\lambda g_{\alpha\beta}\right)$ are the \textit{Christoffel symbols}. Alternatively, one can use the thermodynamic relation $dU = - P dV$, where $U = \rho V$ and $V\propto a^3$. Either way, it is straightforward to obtain the \textit{continuity equation}
\begin{equation}\label{eq:continuityequation}
\dot{\rho} + 3 H (\rho + P ) = 0
\end{equation}
In the standard cosmological model, the medium dominating the energy density in the universe is typically a fluid with equation of state $w\equiv P/\rho$. In this case, the solution to Eq.(\ref{eq:continuityequation}) scales as 
\begin{equation}\label{eq:continuityequationsol}
\rho(a) \propto a^{-3(1+w)}
\end{equation}
\indent Let us now relate the fluid matter to the evolution of $a$ in the FRW universe. To his end, we use Einstein equations
\begin{equation}\label{eq:EinsteinEquation}
G_{\mu\nu} = 8 \pi G T_{\mu\nu}
\end{equation}
where $G$ is Newton's constant and $G_{\mu\nu}= R_{\mu\nu} - \frac{1}{2}R g_{\mu\nu}$
is the Einstein tensor. Here
$R_{\mu\nu} = \partial_\lambda \Gamma^\lambda_{\mu\nu}
- \partial_\nu \Gamma^\lambda_{\mu\lambda} + \Gamma^\lambda_{\lambda\rho} \Gamma^\lambda_{\mu\nu} -\Gamma^\rho_{\mu\rho} \Gamma^\lambda_{\nu\rho}$
is the Ricci tensor and $R= R^\mu_\mu = g^{\mu\nu}R_{\mu\nu}$ is the Ricci scalar. Note, the Einstein tensor is a measure of the spacetime curvature in the universe and therefore Einstein's equation relate curvature to matter. In the case of the FRW metric in Eq.(\ref{eq:FRWmetric}), the only non-zero components of the left hand side of Eq.(\ref{eq:EinsteinEquation}) are 
\begin{equation}\label{eq:FRWGmunu1}
G^0_0 = 3\left(\frac{\dot{a}}{a}\right)^2
\end{equation}
\begin{equation}\label{eq:FRWGmunu2}
G^i_j = \left[2 \frac{\ddot{a}}{a} + \left(\frac{\dot{a}}{a}\right)^2\right]\delta^i_j
\end{equation}
Combining these two equations with Eq.(\ref{eq:Tmunumatrix}) then leads to the \textit{Friedmann equations}
\begin{equation}\label{eq:Friedmann1}
\left(\frac{\dot{a}}{a}\right)^2 = \frac{8\pi G}{3}\rho
\end{equation}
\begin{equation}\label{eq:Friedmann2}
\frac{\ddot{a}}{a} = - \frac{4 \pi G}{3}\left(\rho + 3 P\right)
\end{equation}
Note that the condition for accelerated expansion is $\rho + 3P<0$, or equivalently $w<-1/3$. It is convenient to rewrite the Friedmann equations in terms of $H$ and $\dot{H}$ as follows
\begin{equation}\label{eq:Friedmann3}
\rho = 3 M_p^2 H^2
\end{equation}
\begin{equation}\label{eq:Friedmann4}
P = -M_p^2\left(2\dot{H} + 3H^2\right)
\end{equation}
where $M_p = (8\pi G)^{-1/2} \simeq 2.4 \times 10^{18}$GeV is the \textit{reduced Planck Mass}. Assuming as before that $P = w \rho$ (note since we have three unknowns, $\rho$, $P$ and $a$, an equation of state must be provided in order to close the system of equations) the solution for the scale factor is
\begin{equation}\label{eq:scalefactor}
a(t) = a_0 t^{\frac{2}{3(1+w)}}
\end{equation}
and we thus have (via $\rho \propto H^2 \propto (\dot{a}/a)^2$)
\begin{equation}\label{eq:scalefactor}
\rho(a) = \rho_0 a^{-3(1+w)}
\end{equation}
where $a_0$ and $\rho_0$ are integration constants fixed by the choice of initial conditions. This is of course the same result we found in Eq.(\ref{eq:continuityequationsol}) using conservation of energy. In table \ref{tab:scalefactors}, we show the evolution of the scale factor and the energy density for different values of $w$. Note, $w=-1$ corresponds to a vacuum energy dominated universe (which, as we will see, drives a period of inflation), while $w=1/3$ and $w=0$ correspond to radiation and matter, respectively.
\begin{table}[h!]
\begin{center}
\begin{tabular}{|c|c|c|c|c|c|}
\hline
& $w=-1$ & $w=-\frac{2}{3}$ & $w=\frac{1}{3}$ & $w=0$ \\\hline
$a(t)\propto t^{\frac{2}{3(1+w)}}$& $e^{H t}$ & $t^2$ & $t^{1/2}$ & $t^{2/3}$ \\\hline
$\rho(a) \propto a^{-3(1+w)}$ & Constant & $a^{-1}$ & $a^{-4}$ & $a^{-3}$ \\\hline
\end{tabular}
\end{center}
\caption{The scale factor and energy density for different values of $w$.} \label{tab:scalefactors}
\end{table} 
\\
\indent The equation of state of a static (frustrated) domain wall network is $w= -2/3$ \cite{KolbTurner}. We can then see why a domain wall dominated universe cannot be modeled by a perfect fluid: since the relation between $w$ and the scalar (longitudinal) speed of sound $c_s$ for a perfect fluid is (see Appendix \ref{app:NewtonianPert})
\begin{equation}
c_s^2=w
\end{equation}
a negative value for $w$ would imply the existence of unstable solutions at small scales. Instead, the domain wall network is best modeled by an inflating relativistic elastic solid \cite{SDM} with $w=-2/3$ and an effective rigidity \cite{Rigidity} $\mu=(4/15)\bar{\rho}$, where the value for $\mu$ follows from explicitly computing the change in the energy density due to small shears in an initially isotropic distribution of domain walls. Note, from a symmetry perspective, the choice of a solid seems sensible inasmuch as domain wall networks and solids share the same symmetries -- spatial diffeomorphisms are spontaneously broken in both cases. The special feature of this solid is that it has support for an extra transverse degree of freedom. In general, the speeds of sound for the solid are (see Chapter \ref{ch:ElasticBodies})
\begin{equation}
c_s^2= \frac{d P}{d\rho} + \frac{4}{3} c_v^2
\end{equation}
for scalar modes and
\begin{equation}
c_v^2=\frac{\mu}{\bar{\rho}+\bar{P}}
\end{equation}
for vector (transverse) modes. For $w=-2/3$ (note, $dP/d\rho = w$ for a constant equation of state) and $\mu=4/15$ we have $c_s^2= 2/5>0$ and $c_v^2= 4/5>0$, so the small-scale instabilities disappear.

\section{Inflationary Cosmology}
\label{sec:InfCosmo}
An expanding universe and a finite speed of light imply the existence of horizons. One can define the \emph{particle horizon} as the maximum distance from which a particle of light could have traveled to an observer in the age of the universe. If two particles are separated by a distance larger than the particle horizon, then the particles could never have communicated.
\\
\indent This poses an issue for the standard Big Bang cosmology: the temperature of the CMB is uniform to better than one part in 100,000, yet, for FRW cosmologies, this homogeneity spans scales that are much larger than the particle horizon at the time when the Cosmic Microwave Background (CMB) was formed. If the CMB is made of many disconnected patches that were never in causal contact with each other then why is the CMB temperature uniform across these patches? This is the so-called \emph{horizon problem}. The only way to explain the smoothness of the CMB temperature in the Big Bang theory is by invoking an extreme fine-tuning of the initial conditions of the universe. Though technically not a wrong option, it is extremely unappealing to most physicist who seek a more dynamical explanation.
\\
\indent The horizon problem is beautifully addressed by \emph{inflation}, which postulates the existence of a period of accelerated expansion in the very early universe, before it became dense and hot. As mentioned in the Introduction, inflation also explains why the universe is flat, why we don't see magnetic monopoles, and naturally provides the correct initial conditions for the formation of large-scale structure, like galaxies and galaxy clusters, in our universe. The latter is due to quantum fluctuations in the inflationary matter, which inevitably get magnified to cosmic scales and become the seeds for growth of structure via gravitational instability. Moreover, the small temperature anisotropies observed in the CMB are believed to have originated from these fluctuation, so by carefully studying the CMB data, we can learn about the spectrum of perturbations during inflation. For a great review on inflation see \cite{Baumann:2009ds}.

\subsection{The Horizon Problem}
Let us see how inflation solves the horizon problem: why is the CMB uniform? To study the propagation of light in an expanding spacetime, like FRW, it turns out to be very convenient to switch from physical time $t$ to \textit{conformal time} $\tau$, using $$d\tau = \frac{dt}{a(t)}$$
The reason is that now the propagation of light in FRW is the same as in Minkowski space. Indeed, focusing on the radial component only (because of isotropy) the metric is just
\begin{equation}\label{eq:FRWconformal}
ds^2 = a(\tau)^2\left[d\tau^2 - dr^2\right]
\end{equation}
and we see that null geodesics $ds^2=0 \Rightarrow \Delta r = \pm \Delta \tau$ are simply given by straight lines in a spacetime diagram (note, the lines would be curved if we used $t$ instead).
\\
\indent 
Let us consider the particle horizon. The particle horizon, which limits the distance that a past event can be seen by an observer (and which should not be confused with the ``event horizon'', which concerns 
future events) is given by 
\begin{equation}\label{eq:particlehorizon}
r_{ph}(\tau) = \tau - \tau_i = \int_{t_i}^{t}\frac{dt}{a(t)} = \int_{a_i}^{a}\frac{da}{a\dot{a}} = \int_{\ln a_i}^{\ln a} \mathcal{H}^{-1}d\ln a
\end{equation}
where $\mathcal{H}^{-1} = (aH)^{-1}$ is the \textit{comoving Hubble radius} and $t_i\equiv 0$ is the Big Bang ``singularity'', that is, the moment in time (not space!) when the Big Bang started. The particle horizon then tells you what is the largest comoving distance from which an observer at some time $t$ will be able to receive light signals. The comoving Hubble radius, on the other hand, only concerns the distance from which an observer will be able to receive light signals in that \textit{current} expansion time.
\\ 
\indent For a fluid with $w=P/\rho$ we have 
\begin{equation}\label{eq:ComovingHubbleRadius}
\mathcal{H}^{-1} = H_0^{-1} a^{\frac{1}{2}(1+3w)}
\end{equation}
and since ordinary matter satisfies the strong energy condition, $1+3w>0$, the comoving Hubble radius \textit{increases} with time. Thus, the integral in $r_{ph}$ is dominated at late times. Indeed, solving the integral in the limit where $a_i\rightarrow 0$ and with $w>-1/3$, gives 
\begin{equation}\label{eq:particlehorizonsol}
r_{ph}(a)= \frac{2 H_0^{-1}}{1+ 3w}\left[a^{\frac{1}{2}(1+3w)} - a_i^{\frac{1}{2}(1+3w)}\right] \simeq \frac{2 H_0^{-1}}{1+ 3w}a^{\frac{1}{2}(1+3w)} = \frac{2}{1+ 3w}\mathcal{H}^{-1} 
\end{equation}
Note that in the standard cosmology the particle horizon is roughly the same as the Hubble radius and it is customary (though somewhat misleading) to refer to either one as the ``horizon''.
\\
\indent We have found that the particle horizon $r_{ph}$ in an FRW universe with ordinary matter is finite (defining $\tau_i\equiv \frac{2}{1+ 3w}a_i^{\frac{1}{2}}(1+3w)$, we get $\tau_i\rightarrow0$ as $a_i\rightarrow 0$) and that it grows as $\tau$. Having a Hubble sphere that only grows is problematic in a universe where recombination happened only 300,000 years after the Big Bang. In this case, the particle horizon of an observer today turns out to be about $10^4$ times larger than the particle horizon of causal observer at the time of recombination, when the CMB was formed, and this means that the CMB is made of about $10^4$ causally disconnected regions. There was not enough time between the singularity and recombination for the regions to be in causal contact. 
\\
\indent A simple way to solve this is to introduce a fluid that violates the strong energy condition, that is, it satisfies $1 + 3w <0$. From Eq.(\ref{eq:ComovingHubbleRadius}), this implies a \textit{shrinking} Hubble sphere since we now have $d(\mathcal{H}^{-1})/dt<0$. Moreover, the singularity ($a_i\rightarrow 0$) is now at 
$$\tau_i= \frac{2}{1+3w}a_i^{\frac{1}{2}(1+3w)}\rightarrow-\infty$$
so there was plenty of (conformal) time between the singularity and recombination for the $10^4$ CMB regions to be in causal contact. Having a shrinking comoving Hubble sphere ($d(a H)^{-1}/dt<0$) is equivalent to a period of inflation since it implies:
\begin{itemize}
\item Accelerated Expansion: $\frac{d}{dt}(a H)^{-1}=\frac{d}{dt}(\dot{a})^{-1} = - \frac{\ddot{a}}{(\dot{a})^2}<0 \Rightarrow \ddot{a}>0$
\item Negative pressure: Eq.(\ref{eq:Friedmann2}) with $\ddot{a}>0$ implies $\rho + 3P<0$
\item Slowly-Varying Hubble: $\frac{d}{dt}(a H)^{-1}<0\Rightarrow \epsilon\equiv -\frac{\dot{H}}{H} <1$ 
\end{itemize}
 Note that $\epsilon<1$ is enough to produce an accelerated expansion, but for this expansion to be exponential one must also require that $\epsilon\ll1$. Then, one gets $H=\dot{a}/a\sim const.$ and $a(t) = e^{Ht}$. We also see that, since $\rho\propto H^2 = const.$, inflation can be generated with a source of matter whose energy density does not change rapidly.
\\
\indent So, how can we guaranty that inflation solves the horizon problem? First, consider two particles separated by some distance $\lambda$. Then, if $\lambda> (a H)^{-1}$, the particles cannot communicate at the current expansion time, whereas if $\lambda>r_{ph}$, the particles could have never communicated in the history of the universe. Therefore, one conservative solution to the horizon problem is to demand that the observable universe was inside the  Hubble radius at the beginning of inflation, that is
\begin{equation}
(a_0H_0)^{-1} < (a_i H_i)^{-1}
\end{equation}
where the subscript 0 labels quantities today and the subscript $i$ labels quantities at the start of inflation. 

\subsection{Slow-roll Inflation}
We have seen that the condition $\epsilon<1$ leads to a rapidly expanding universe. However, for this expansion to last a non-trivial amount of time (around 60 e-folds) we also need that
$$\eta \equiv \frac{\dot{\epsilon}}{H \epsilon} < 1$$
It turns out that a simple scalar field provides the necessary micro-physics so that both of these conditions are satisfied. The contribution to the stress-energy tensor from the inflaton is 
\begin{equation}
T_{\mu\nu} = \partial_\mu\phi\partial_\nu \phi - g_{\mu\nu}\left(\frac{1}{2}g^{\alpha\beta}\partial_\alpha\phi\partial_\beta\phi - V(\phi)\right)
\end{equation}
where $V(\phi)$ is the field's potential. At the background level the inflaton is homogeneous, and inserting $\phi(t)$ in the above equation gives
\begin{align}
\rho_\phi &= T^0_0 = \frac{1}{2}\dot{\phi}^2 + V(\phi) \\
P_\phi &= -\frac{1}{3}T^i_i = \frac{1}{2}\dot{\phi}^2 - V(\phi) 
\end{align}
Note, the first equation is just kinetic plus potential energy whereas the second equation is  kinetic minus potential energy. Plugging these two equations into the Friedman relations, Eq.(\ref{eq:Friedmann3}) and Eq.(\ref{eq:Friedmann4}), we get
\begin{align}\label{eq:Hscalarfield}
H^2 &= \frac{\rho_\phi}{3 M_p^2} = \frac{1}{3M_p^2}\left[\frac{1}{2}\dot{\phi}^2 + V(\phi)\right] \\\label{eq:Hdotscalarfield}
\dot{H} &= - \frac{\rho_\phi + P_\phi}{2 M_p^2} = -\frac{1}{2}\frac{\dot{\phi}^2}{M_p^2}
\end{align}
Thus for inflation to occur we must have 
\begin{equation}\label{eq:epsilon}
\epsilon \equiv  -\frac{\dot{H}}{H^2} = \frac{\frac{1}{2}\dot{\phi}^2}{M_p^2 H^2}<1
\end{equation}
which tells us that the kinetic energy must be smaller than the potential. Taking the time derivative of Eq.(\ref{eq:Hscalarfield}) and using Eq.(\ref{eq:Hdotscalarfield}) we obtain the evolution equation for a background scalar field in an expanding universe:
\begin{equation}\label{eq:phiEqOfMotion}
\ddot{\phi} - 3 H \dot{\phi} + V' = 0 
\end{equation}
where here (and from now on) the prime on $V$ stands for $dV/d\phi$ (not $dV/d\tau$). This equation says that the acceleration of $\phi$ (first term) is set by the friction term coming from the universe's expansion (second term) and the force induced by the potential (last term). For inflation to last long, however, the acceleration term must be small, that is, one must have $\delta\equiv -\ddot{\phi}/H\phi<1$. Thus, if $\epsilon, \delta \ll 1$ we also have 
\begin{equation}
\eta \equiv \frac{\dot{\epsilon}}{\epsilon H} = 2(\epsilon - \delta) \ll 1 
\end{equation}
In the slow roll approximation
\begin{align}
\epsilon &\equiv \frac{\frac{1}{2}\dot{\phi}^2}{2 M_p^2}\ll1 \quad\Rightarrow\quad H^2 \simeq \frac{V}{3 M_p^2}\\
|\delta| &\equiv \frac{|\ddot{\phi}|}{H |\phi|}\ll 1 \quad \Rightarrow \quad 3 H \dot{\phi} \simeq -V'
\end{align} 
Defining the following \textit{slow roll parameters} 
\begin{align}\label{eq:epsilonV}
\epsilon_V &\equiv \frac{M_p^2}{2}\left(\frac{V'}{V}\right)^2 \simeq \epsilon\\
\eta_V &\equiv M_p^2\frac{V''}{V} \simeq 2 \epsilon - \frac{1}{2}\eta 
\end{align}
we then have successful inflation if $\epsilon_V,\eta_V\ll1$. This approximation is very convenient since we can now construct models of inflation by simply picking ``flat'' potentials that gives $\epsilon_V,\eta_V\ll1$.
\\
\indent Let us close this section with the simplest example of an inflation model: ``chaotic inflation''. In this model, inflation is driven by the mass term 
$$V(\phi) = \frac{1}{2}m^2\phi^2$$
where $m$ is the inflaton mass. The slow roll parameters are thus
\begin{equation}
\epsilon_V(\phi) =\eta_V(\phi) = 2\left(\frac{M_p}{\phi}\right)^2
\end{equation} 
and to have $\epsilon_V,\eta_V<1$ we must consider superplanckian field excursions, $\phi>\sqrt{2}M_p$. The length of the inflationary period is typically measured in ``e-folds'':
\begin{equation}
\N = \int_{a_{i}}^{a_e}d\ln a = \int_{t_i}^{t_e}H dt \simeq \int_{\phi_i}^{\phi_e}\frac{1}{\sqrt{2 \epsilon_V}}\frac{|d\phi|}{M_p} 
\end{equation}
where we have used $H dt = H d\phi/\dot{\phi}$ with $\dot{\phi} \simeq \sqrt{2  \epsilon_V}M_p$ (valid only in the slow roll approximation, Eq.(\ref{eq:epsilonV})). For the chaotic inflation model we then have 
$$\N(\phi) = \frac{\phi^2}{4 M_p^2} - \frac{1}{2}$$
It then turns out that for inflation to solve the horizon problem one must have $\N\gtrsim 60$, so the fluctuations observed in the CMB are created when $\phi \sim 15 M_p$.

\subsection{Primordial Perturbations}
We have already mentioned that quantum fluctuations during inflation give rise to the primordial perturbations which seed all of the large-scale structures we see around us. In this section we briefly review how this happens.
\\
\indent The basic idea is simple. We just showed that the duration of the inflationary period is controlled by the inflaton field, so $\phi(t)$ acts as a ``clock'' which measures the time to the end of inflation. After quantizing the field, the uncertainty principle says that this clock will inevitably have an variance, so the inflaton field will have space-dependent fluctuations $\delta\phi(x,t) = \phi(x,t) - \bar{\phi}(t)$ and locally, inflation will end at different times $\delta t(x)\sim \delta \phi(x)/\dot{\phi}$. Working in the so-called ``flat gauge'', where all of the scalar perturbations are captured by $\delta\phi$, this local time difference will produce curvature perturbations 
\begin{equation}\label{eq:deltaphitoR}
\CR \sim H\delta t \sim \frac{H}{\dot{\phi}}\delta\phi= \frac{\CH}{\phi'} \delta\phi
\end{equation}
These curvature perturbations then induce density perturbations $\delta \rho (x)$ (seeding the gravitational instabilities which pull dark matter together and create structure in the universe) and ultimately, the temperature anisotropies $\delta T(x)$ observed in the CMB.
\\
\indent From Eq.(\ref{eq:deltaphitoR}), we see that the variance of the curvature perturbations depends on the variance of $\phi$ as follows (in Fourier space)
\begin{equation}\label{eq:varianceofR}
\left\langle|\CR_k|^2\right\rangle = \left(\frac{\CH}{\phi'}\right)^2\left\langle|\delta\phi_k|^2\right\rangle
\end{equation}
To compute $\left\langle|\delta\phi_k|^2\right\rangle$ we first need to solve for the classical dynamics of $\phi$. This can be derived from the inflaton action 
\begin{equation}\label{eq:ClassicalAction}
S = \int d\tau dx^3\sqrt{-g}\left[\frac{1}{2}g^{\mu\nu}\partial_\mu\phi\partial_\nu \phi - V(\phi)\right]
\end{equation}
where $g\equiv \det(g_{\mu\nu})$. Defining $f(x,t)\equiv a(\tau)\delta\phi(x,t)$, working in the flat gauge, and applying the slow-roll approximation, the quadratic action becomes \cite{MukhanovBook, Baumann:2009ds}
\begin{equation}\label{eq:ClassicalAction}
S_2 \simeq \frac{1}{2}\int d\tau dx^3\left[\left(f'\right)^2 - \left(\nabla f\right)^2 + \frac{a''}{a}f^2\right]
\end{equation}
and from the Euler-Lagrangian equations one can derive the well-known \textit{Mukhanov-Sasaki equations} (MS)
\begin{equation}\label{eq:MSeqs}
f''_k + \left(k^2 - \frac{a''}{a}\right)f_k =0
\end{equation}
In the case of slow-roll inflation (where de Sitter is a good approximation for the background expansion), $a''/a\simeq 2 \CH^2 \simeq 2/\tau^2$ and one can solve the MS equation to get the so-called ``Bunch-Davies'' solution:
\begin{equation}\label{eq:MSsol}
f_k(t) = \frac{e^{-i k\tau}}{\sqrt{2k}}\left(1 + \frac{i}{k\tau}\right)
\end{equation}
We can now compute the statistics of the quantum operator $\hat{f}_k = f_k(\tau) \hat{a}_k + f_k^*(\tau) \hat{a}_k^{\dagger}$, where $\hat{a}_k$ and $\hat{a}^{\dagger}_k$ are the usual lowering and raising operators (which satisfy the commutation relation $[\hat{a}_k,\hat{a}^{\dagger}_k]=\delta(k + k')$ and define the vacuum state via 
$\hat{a}_k |0\rangle =0$). It is straightforward to show that the variance of the operator in position space is
\begin{align*}
\langle |\hat{f}|^2 \rangle &\equiv \langle 0| \hat{f}^{\dagger}(\tau,0)\hat{f}(\tau,0)|0 \rangle \\
&=\int \frac{d^3 k}{(2\pi)^{3/2}} \frac{d^3k'}{(2\pi)^{3/2}} f_k(\tau)f^*_{k'}(\tau) \langle 0| [\hat{a}_k,\hat{a}_{k'}^{\dagger}]|0 \rangle\\
&= \int d\ln k \mathcal{P}_f(k,\tau)
\end{align*}
where we have defined the \textit{dimensionless power spectrum} 
\begin{equation}
\mathcal{P}_f(k,\tau) \equiv \frac{k^3}{2 \pi^2}|f_k(\tau)|^2
\end{equation}
Using the Bunch-Davies solution in Eq.(\ref{eq:MSsol}), we can then obtain an expression for the inflaton spectrum 
\begin{equation}\label{eq:PhiSpectrum}
\mathcal{P}_\phi(k,\tau) = \frac{\mathcal{P}_f(k,\tau)}{a^2} = \left(\frac{H}{2\pi}\right)^2\left(1 + \left(\frac{k}{a H}\right)^2 \right) \simeq \left(\frac{H}{2\pi}\right)^2
\end{equation}
where in the last equality we took the superhorizon limit $k\ll a H$. Note, this result means that the quantum fluctuations of the inflaton field during a de Sitter phase are of the same order as the Hubble scale at horizon exit. Note also that during inflation, and in comoving coordinates, the wavelengths of fluctuations are constant while the Hubble radius shrinks, effectively stretching the fluctuations to superhorizon scales. At these large scales, the fluctuations stop being quantum mechanical in nature and ``turn into'' fluctuations in a stochastic field.
\\
\indent Using Eq.(\ref{eq:PhiSpectrum}), Eq.(\ref{eq:varianceofR}) and Eq.(\ref{eq:epsilon}), we can finally obtain the curvature power spectrum at horizon-exit:
\begin{equation}\label{eq:primordialscalarinf}
\mathcal{P}_{\CR}(k,\tau) =  \left(\frac{H^2}{8 \pi^2 \epsilon M_p^2}\right)_{k=a H} \equiv A_s \left(\frac{k}{k_*}\right)^{n_s -1}
\end{equation}
Experimentally, we have measured the \textit{amplitude} to be \cite{Planck} $A_s \sim 10^{-9} $ and the \textit{spectral index} to be $n_s\simeq 0.96$ at the pivot scale $k_* \simeq 0.05$Mpc$^{-1}$. These are the \textit{late-time observables }that allow us to learn about the early universe during inflation. Note that we only need to evaluate the spectrum at horizon exit because (as we will see in Chapter \ref{ch:CosmoPert}) the curvature perturbations remain constant outside the horizon in the standard cosmology without a domain wall era. Lastly, a similar analysis as the one above can be applied to tensor perturbations, $h$ (we defined $h$ more precisely in Chapter \ref{ch:CosmoPert}). The final result is
\begin{equation}\label{eq:primordialtensorinf}
\mathcal{P}_{h}(k,\tau) =  \left(\frac{2H^2}{\pi^2 M_p^2}\right)_{k=a H}
\end{equation}

\section{Domain Wall Networks}\label{sec:Review}
Formally, the presence of a particular defect after spontaneous symmetry breaking is determined by the topology of the \emph{vacuum manifold} $\mathcal{M}=G/H$ of the theory or model in question, where $G$ and $H$ are the symmetry groups before and after the breaking, respectively \cite{Kibble:1976sj}. If the $n^{th}$ homotopy group of the vacuum manifold is nontrivial ($\pi_{n}(\mathcal{M})\neq \mathbb{I}$) then there exist inequivalent mappings between $n$-dimensional spheres ($S^{n}$) and $\mathcal{M}$, meaning that there exist closed $n$-dimensional surfaces, in the field space where the vacuum manifold lives, which pass through a given point but that cannot be continuously deformed into each other. In $3$ spatial dimensions, domain walls are sheet-like surfaces which form if $\pi_{0}(\mathcal{M})\neq \mathbb{I}$, or in other words, if the vacuum manifold is made of disconnected points. Similarly, strings are line-like defects which form if $\mathcal{M}$ contains unshrinkable loops ($\pi_{1}(\mathcal{M})\neq \mathbb{I}$), and monopoles are point-like defects which form if $\mathcal{M}$ contains unshrinkable surfaces ($\pi_{2}(\mathcal{M})\neq \mathbb{I}$). 
\\
\indent Among these defects, domain walls are the simplest since they arise if the potential of a field has a global discrete symmetry that is spontaneously broken by the vacuum. In our cosmological context, the universe after the phase transition divides into domains, each populated at random by one of the available vacua. We now give some examples of models which can produce domain walls in the universe and we also discuss what we know about their evolution.

\subsection{$Z_2$ (double-well model)}
\begin{figure}[t]
\center{
\includegraphics[scale=0.4]{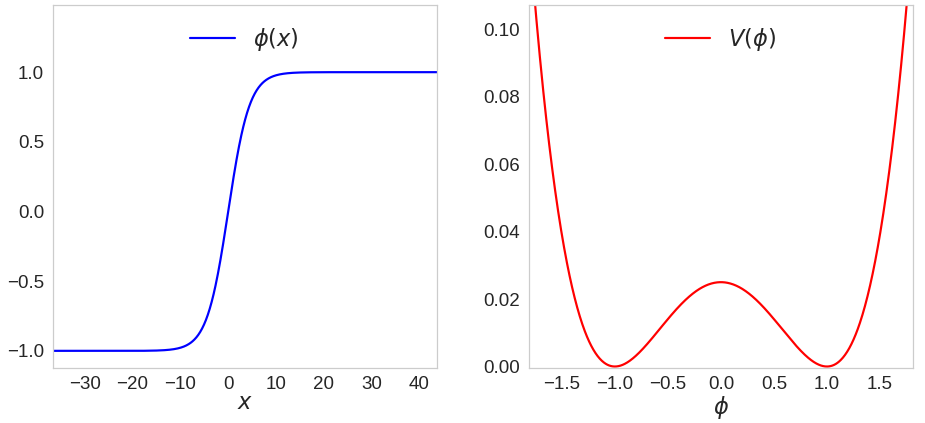}
\caption{Plot of the $Z_2$-wall solution $\phi(x) = v \tanh(\sqrt{\lambda/2}v x)$ and the potential $V(\phi)= \lambda(\phi^2-v^2)^2/4$ (here, $v=1$, $\lambda=0.1$ so $\delta_w\simeq 4.5/v$).\label{fig:DWsol}}}
\end{figure}

The simplest example of a model producing domain walls is the double well potential in $1$ spatial dimension (the walls are also called kinks in this case)
\begin{equation}\label{eq:Z2potential}
V(\phi)=\frac{\lambda}{4}\left( \phi^2 -v^2 \right)^2
\end{equation}
where $\phi(x)$ is a real scalar field. At $\phi=0$, the field is in the \emph{false} vacuum energy state, $V_0 =m^2v^2/4$, where $m\equiv \sqrt{\lambda} v$ is the mass parameter of $\phi$. The $Z_2$ symmetry of the potential is then broken by the \emph{true} vacuum when the field rolls down the potential to either $\phi=\pm v$, where the potential vanishes. Since $\pi_0(Z_2/\mathbb{I})=Z_2$, the phase transition gives rise to domain walls.
\\
\indent The Lagrangian of the system is $\mathcal{L} = \frac{1}{2}(\partial_\mu\phi)^2 - V(\phi)$ and the equation of motion, $\partial^2\phi/\partial x^2 - \partial V/\partial\phi =0$, is then 
\begin{equation}\label{eq:DWEoM}
-\frac{\partial^2\phi}{\partial x^2} + \lambda \phi(\phi^2 -v^2) =0
\end{equation}
The solution to this equation is $\phi(x)=v \tanh(x/\delta_w)$, where $\delta_w = (\lambda/2)^{-1/2}v^{-1}=\mathcal{O}(1/m)$ characterizes the ``thickness'' of the wall (see Fig.(\ref{fig:DWsol})).
Note, the solution is centered about $x=0$ and interpolates the value of $\phi$ from $-v$ at $x=-\infty$ to $v$ at $x=+\infty$. Moreover, these boundary conditions, combined with the non-trivial topology of the vacuum, guarantee that potential energy is stored somewhere between the boundaries, where the field is outside the vacuum manifold. This energetic region of thickness $\delta_w$ is the domain wall and because it is time-independent, it is classically stable.
\\
\indent This stability can also be understood as being the result of a topological conservation law where each ``defect state'' has a conserved quantum number associated with it. The conserved topological current follows from the discrete symmetry of the Lagrangian, in contrast with the currents that are conserved by Noether's theorem in the case of continuous symmetries. For this 1D model, the topological current is 
\begin{equation}
J^\mu = \epsilon^{\mu\nu} \partial_\nu \phi
\end{equation}
where $\epsilon^{\mu\nu}$ is the Levi-Civita tensor. Note, this current is trivially conserved because we are taking the divergence of an anti-symmetric tensor. The conserved charge associated with this current is 
\begin{equation}
Q = \int dx J^0 = \phi(\infty) - \phi(-\infty) =2 v 
\end{equation}
and since it is non-zero the configuration is stable.
\\
\indent The surface energy --or tension -- $\sigma$ of this ``$Z_2$ wall'', is given by the balance between a gradient term of order $\delta_w (\partial_x \phi)^2\sim \delta_w^{-1}v^2$ and a potential energy term of order $\delta_w V_0$. Minimizing the gradient energy favors a large wall thickness, whereas minimizing the potential energy favors a small thickness. Thus, to minimize the surface energy, the two contributions must balance each other. Setting the gradient and potential energies equal to each other, we get $\delta_w\sim1/m$ (which is of course the same result we got directly from the domain wall solution), and a  wall tension 
\begin{equation}
\sigma\sim m v^2
\end{equation}
In the case of the $Z_2$-wall, we can actually compute the tension exactly from the stress-tensor $T_{\mu\nu} = \partial_\mu\phi \partial_\nu \phi - g_{\mu\nu} \mathcal{L}$. This gives 
\begin{align}\label{eq:Tmunuwalls}
T^\mu_\nu &= \left(\frac{\partial\phi}{\partial x}\right)^2\text{diag}(1,0,1,1)\\
&=\frac{1}{2}\lambda v^4\left[\cosh\left((\lambda/2)^{1/2} v x\right)\right]\times \text{diag}(1,0,1,1)
\end{align}
where the first equality follows from integrating once the equation of motion to get $\frac{1}{2}(\partial \phi/\partial x)^2 - V(\phi) =0$, and the second equality just uses the wall solution $\phi(x) = v \tanh(\sqrt{\lambda/2}v x)$. Note that the surface energy is also equal to the tension along the $y$ and $z$ directions. The surface energy density can now be computed exactly as
\begin{equation}
\sigma = \int T_0^0 dx = \frac{2\sqrt{2}}{3}\sqrt{\lambda}v^3
\end{equation}
In 3D, the wall is localized at $x=0$ and the field configuration is Lorentz invariant with respect to boosts in the $yz$-plane; so only motion that is transverse (in the $x$-direction) is relevant.
\\
\indent Let us assume that one of these $Z_2$ walls was subdominantly produced (to say, radiation) in our universe with an initial characteristic curvature radius of order $R$, in real space, smaller than the horizon. In this case, the tension in the wall will produce a force per unit area $f\sim \sigma/R$ that will stretch and flatten the wall at relativistic speeds, and though particle scattering could damp this motion, the walls will dominate the energy density of the universe in a time-scale no longer than $\sim(G\sigma)^{-1}$, where $G$ is Newton's constant (we show this in Section \ref{ssec:NetworkEvolution}, when we discuss the evolution of domain wall networks). 
\\
\indent An interesting feature of domain walls is that they are gravitationally repulsive. This can be seen immediately in the Newtonian limit of Einstein equations for a static wall lying in the $yz$-plane. Assuming that the wall is ideally thin, so that $T_\mu^\nu = \sigma \delta(x)\text{diag}(1,0,1,1)$, the gravitational potential $\Phi$ in this limit satisfies
\begin{equation}
\label{eq:EMT}
\nabla^2\Phi=4 \pi G (T_0^0-T_i^i)=-4\pi G \sigma \delta(x)<0
\end{equation}
which evidently implies that the wall is a source of gravitationally repulsive force.
\\
\indent The relativistic treatment is a bit more subtle. Consider a universe with only one domain wall and nothing else. Naively, one might envision a domain wall as a static object with planar and reflectional symmetry. Though this was intuitive in the Newtonian limit, this simple interpretation is skewed when one considers a fully general relativistic approach. In fact, it can be shown that no static solutions of Einstein's equation exist with the planar and reflectional symmetry and an energy momentum tensor as in Eq.(\ref{eq:Tmunuwalls}). A time dependent solution, however, was found by Vilenkin \cite{Vilenkin:1984hy} and Ipser $\&$ Sikivie \cite{PhysRevD.30.712}. The solution they found is 
\begin{equation}
ds^2=(1-\kappa|x|)^2dt^2-dx^2-(1-\kappa|x|)^2e^{2\kappa t}(dy^2 + dz^2).
\end{equation}
where $\kappa=2\pi G \sigma$. In summary, it combines a $(1+1)$-dimensional Rindler metric in the $(x,t)$ components and a (2+1)-dimensional de Sitter metric on constant $x$ hypersurfaces. Therefore, a test particle would be perceived by an observer at $x=0$ as moving away from the wall with an acceleration $a=\kappa$ until it reaches the event horizon at $|x|=\kappa^{-1}$. In other words, the walls are gravitationally repulsive, as was expected form the Newtonian limit result. The $yz$-plane has a de Sitter horizon of radius $\kappa^{-1}$ as well, which is also consistent with the intuition that the residual false-vacuum energy inside the wall would drive inflation inside the wall. Lastly, transforming the metric to Minkowski form on either side of the wall, would reveal that an inertial observer outside the wall region would see itself enclosed by a sphere with an outward acceleration of $a=\kappa$.
The metric in a universe with one wall is locally flat everywhere, except on the wall \cite{PhysRevD.30.712}.

\subsection{$Z_N$ (axion models)} \label{ssec:axions}
Axion models typically involve the $U(1)$ symmetric potential of a complex scalar field $\Phi$, that gets spontaneously broken by the vacuum, producing axion strings and massless Goldstone bosons (the axions) in the process. The $U(1)$ symmetry is then explicitly broken to its $Z_N$ subgroup by non-perturbative (instanton) effects from a strong gauge (QCD or hidden) sector. The $Z_N$ symmetry implies an $N$ fold degeneracy of the vacuum manifold, corresponding to equidistant points on the circle at the bottom of the ``Mexican hat'' potential. Therefore, these models can produce a hybrid network of domain walls bounded by strings \cite{AxionsSikivieEtAl}.
\\
\indent Let us give a simple example of a $Z_N$-symmetric potential. Consider a complex scalar field that is invariant under a $Z_N$ symmetry, $\Phi\rightarrow \Phi \exp(i 2\pi k/N)$, where $k$ takes integer values. A  Lagrangian with $N$ vacua is then \cite{PhysRevLett.110.021803}:
\be
\mathcal{L}=|\partial_\mu \Phi|^2 - V(\Phi); \quad\quad V(\Phi)=\lambda v^4\left[\left(\frac{2^{1/2}\Phi}{v}\right)^N-1\right]^2
\ee
where $v$ is an energy scale and $\lambda$ is a constant. Writing $\Phi=2^{-1/2}\phi\exp(i a/v)$ one can show that the potential is minimized when the radial Higgs mode is $\phi=v$ and when the axion-like field $a$ takes the value $$a=v\times\left\lbrace 0, \frac{2\pi}{N}, \frac{4\pi}{N}\cdots,\frac{2\pi(N-1)}{N}\right\rbrace$$ 
At the $\phi=v$ minimum, the effective theory for the axion is
\be
\mathcal{L}_a=\frac{1}{2}\left(\partial_\mu a\right)^2 - 4\lambda v^4\sin^2\left( \frac{Na}{2v}\right)
\ee
The field configuration interpolating between two adjacent minima represents a domain wall solution. The solution for a wall in the $yz$-plane (centered at $x=0$) interpolating the minimum at $a=0$ and the minima at $a=2\pi v/N$, is 
\begin{equation}
a(x) = \frac{4 v}{N}\times \arctan\left(e^{m_a x}\right);
\qquad\quad\frac{da}{dx}= \frac{2 v m_a}{N \cosh\left(m_a x\right)}
\end{equation}
where $m_a = N \sqrt{2\lambda} v$ is the axion mass setting the wall thickness $\delta_w = 2/m_a$ (from balancing gradient and potential energies). The domain wall tension (mass/area) can then be found by integrating the gradient energy
\begin{equation}
\sigma = \int dx \frac{da}{dx} = \frac{8 m v^2}{N^2}
\end{equation}
If $N\geq 3$, the walls can join at strings and form a network which could be long lived, even though annihilation processes take place \cite{AxionsSikivieEtAl}.

\subsection{Approximate $O(N)$}\label{ssec:Kubotani}
A more isotropic network is produced in the following model \cite{ONSymKubotani}
\begin{equation}
V(\chi_1,\chi_2,\cdots, \chi_N) = \frac{\lambda}{4}
\left(\sum_i^N\chi_i^2 -v^2\right)^2 + \epsilon \sum_i^N\chi_i^4\end{equation}
where each $\chi_i$ is a scalar field and where $|\epsilon|\ll\lambda$. This potential has a $O(N)$ symmetry (given by the first term) that is broken spontaneously at some scale $v$, but also explicitly broken by the last term. The $O(N)$ is said to be approximate, and since it is now discrete, domain walls will form. If $\epsilon<0$, there are $2N$ minima: $\chi^{(i)}_{min}=\{0,\cdots,0,\pm v,0,\cdots,0\}$ for each $\chi_i$. On the other hand, if $\epsilon>0$, there are $2^N$ minima at
\be
\chi_{min}=\left\lbrace\pm \frac{v}{\sqrt{N}},\cdots,\pm \frac{v}{\sqrt{N}}\right\rbrace;\qquad v = \frac{m}{\sqrt{\lambda}}
\ee
for all possible sign combinations and where $m$ is the mass parameter for each $\chi_i$. Since there are lots of minima that regions in space can settle at, pair annihilation processes of walls are expected to be suppressed. Moreover, the walls can meet at junctions due to the non-abelian nature of the underlying symmetry, so the vacuum configuration is rich enough to form an isotropic network that can initially frustrate. We will revisit this model in the context of hybrid inflation in Chapter \ref{ch:Model}.

\subsection{Network Evolution}\label{ssec:NetworkEvolution}
Let us explore what happens when a network with a characteristic scale much smaller than the Hubble volume is produced. Let us ignore the universe's expansion for now and use percolation theory \cite{Stauffer:1978kr} to estimate how many domains are left when all neighboring domains with same vacua merge and the system stabilizes. Percolation theory is a mathematical model that describes the growth of ``clumps'' or \emph{clusters} in a random environment typically represented by a lattice. It's range of applications is broad since the results rely mostly on combinatorics rather than the specific nature of the clusters. It is used, for example, to model the flow of fluids through a porous medium, or the propagation of fire through a forest.
\\
\indent In our cosmological context, percolation theory models the growth of the domains inside a static Hubble patch. The initial Hubble patch is represented by a cubic lattice of length $H_i^{-1}$ and lattice spacing $L_0$, assumed to be much smaller than $H^{-1}_i$. The stable domain walls then live between cells separated by different types of vacua. As we will see in Chapter \ref{ch:Model}, the model we present can lead to these initial conditions. Assuming that each defect field randomly rolls down to one of the available vacua in a given spatial region, each vacuum has a probability $p=1/n_v$ to populate a given cell, where $n_v$ is the number of distinct vacua in the model. Given $p$, using percolation theory we can find the average size of the domains once all neighboring cells with the same vacua have merged.
\\
\indent The fundamental result from percolation theory is the existence of a percolation threshold $p_c$, whose value depends on the choice of lattice. If a particular vacuum has $p\geq p_c$, then the domains occupied by this vacuum will always percolate through the lattice and the system will evolve towards a universe with only one ``infinite'' wall of size $H_i^{-1}$. On the other hand, if $p<p_c$, the domains grow but remain smaller than $H_i^{-1}$ after the system stabilizes. Furthermore, numerical analysis of series data \cite{Flammang} show that for $p<p_c$, the average number of domains with volumes $s$ times larger than the initial cubic cell (an $s$-cluster), decreases exponentially as $e^{-s}$. For the simple $Z_2$ model (where $n_v=2$) and using the percolation threshold of a simple cubic lattice \cite{StaufferAharony}, $p_c=0.311$, we have $p=1/2>p_c$, so each Hubble patch will possess one infinite wall.
\\
\indent As we have already discussed, the metric of a universe dominated by one (or a few) domain walls is very inhomogeneous and anisotropic, and the way the universe expands in this scenario is not known. Therefore, any sensible cosmological model with a domain wall era must involve a network that has a large number of walls inside the horizon throughout its entire lifetime. Indeed, at scales much larger than the network's characteristic length scale, we may still use the nice properties of a Friedmann-Robertson-Walker (FRW) universe, which is homogeneous, isotropic and has a computable Hubble radius evolution.  Since the simple $Z_2$ model is not suitable for a domain wall era, a model with a richer vacuum structure, like the one we present in Chapter \ref{ch:Model}, is needed.
\\
\indent Let us now consider how an isotropic and homogeneous domain wall network redshifts as the universe expands. We will assume that the walls are initially subdominant to radiation. We are interested in the evolution of $L(a)$, the typical physical size of a domain as the universe expands. Since $L$ is related to the energy density of the domain wall network by $\rho_{walls}=\sigma/L$, and since $\rho_{walls} \propto a^{-3(1+w_{walls})}$, the evolution of the domain size can be parameterized by the equation of state $w_{walls}$ of the network. In general, the equation of state for a ``domain wall gas'' is given by \cite{KolbTurner, Zeldovich,Gelmini:1988sf,CosmicStrings}
\begin{equation}\label{eq:w}
w_{walls} = -2/3+ v^2_{rms}
\end{equation}
where $v_{rms}$ is the root mean squared velocity of the walls . The $v^2_{rms}$ term follows from the average kinetic energy in the network, whereas, the -2/3 factor follows from treating each domain wall as false-vacuum energy confined to a 2D plane, and then averaging over the whole space (assuming that the network is isotropic). Finding a precise value for $w_{walls}$ is hard since it depends on the speed of the walls, and therefore requires model-dependent information about the interaction between the walls and their surroundings. Nevertheless, we can still restrict the allowed range of values it can take. 
\\
\indent An extreme possibility (albeit a very unlikely one) is that the network remains static during the radiation era. In this case, $w_{walls}\simeq -2/3$ and the network's energy density dilutes as $\rho_{walls}\sim a^{-1}$. The network then quickly dominates the energy density since $\rho_{walls}$ dilutes at a much slower rate than the energy density of radiation (which goes as $a^{-4}$). Note that $L\propto a$, so the walls are just stretched by the Hubble flow and the network is said to be \emph{frustrated} (or static). During radiation domination the Hubble radius increases as $a^{2}$, which is faster than the growth of $L$. Therefore, more and more walls fall inside a Hubble patch as it grows, and as a result, the number of walls inside the horizon will be larger by the time the network comes to dominate.
\\
\indent A more likely possibility is that the domain wall network reaches a \emph{scaling regime}, where the walls disentangle and the network ``rescales'' itself in such a way that the number of walls inside the expanding Hubble patch remains (approximately) \emph{fixed}. This implies that $L(a)\propto H^{-1}\propto t \sim a^{2}$, that is, as fast as allowed by causality ($L$ cannot grow faster than $H^{-1}$). The network's energy density in this case dilutes as $\rho_{walls}\propto \sigma H \propto a^{-2}$, which is still slower than the dilution of radiation. So, in this regime the walls will also come to dominate the energy density. Note that in general we can write
\begin{equation}
\frac{\rho_{walls}}{\rho_{dominant}} \sim \frac{\sigma H}{3 M_p^2 H^2}\sim G\sigma t
\end{equation}
so the walls always end up dominating the energy budget in a time no longer than $(G\sigma)^{-1}$, regardless of its degree of frustration or the equation of state of the dominant component (assuming it is a fluid like radiation or matter). In reality, the precise network evolution of the subdominant network probably falls somewhere between the scaling and frustrated cases. It is then not hard to show that (assuming radiation domination)
\begin{equation}\label{eq:wbound}
-\frac{2}{3}\leq w_{walls} \leq -\frac{1}{3}
\end{equation}
The lower bound corresponds to the maximally frustrated regime while the upper bound corresponds to the scaling regime. Note, if we instead assume matter domination, the upper bound decreases to $w_{walls}=-1/2$ \cite{Friedland:2002qs} (this is consistent with what simulations have found \cite{Avelino:2008ve}, namely $w_{walls}\simeq-13/24$). 
\\
\indent To find more precise values for $w_{wall}$ one must resort to case-by-case results from simulations. For $Z_2$ walls, with only two minima, the simulations performed in \cite{PRS1989} show that a scaling regime is reached and that there is about one large wall per horizon at any given time. A follow up analysis covering more complicated models, like axions with several vacua\footnote{Axion models with only one vacuum form networks made of single walls terminating on strings which quickly disintegrate before they can dominate \cite{AxionsSikivieEtAl}.} (i.e. with more than one bump in the bottom of the Mexican hat) also concludes that the networks scale with Hubble \cite{PRS1990}. Lastly, the most recent state-of-the-art simulations go a step further and suggest that any model with junctions, even under the most ideal conditions, produces networks that eventually reach the scaling regime \cite{Avelino:2005kn,PinaAvelino:2006ia,Avelino:2006xf,Avelino:2008ve, Leite:2011sc}. One could try to find a mechanism which significantly slows down the network's propensity to disentangle (for example, by turning on non-trivial interactions between the wall's field constituents and the surrounding matter or radiation). However, we will simply take the simulation results at face value and assume that, if initially subdominant, the network evolves as in the scaling regime. 
\\
\indent Once the network dominates the energy density, because $w_{walls}<-1/3$, the universe will start to inflate and we expect that any residual wall velocities will rapidly decay with time due to the Hubble friction and the network frustrates. In other words, since all the mechanisms which could spoil frustration are redshifted away, the equation of state reverts back to $w_{walls}=-2/3$, so that $L\sim a$, $a\sim t^2$ and thus a domain wall inflationary period proceeds. Though frustration during domain wall dominance has not been proven, it is plausible and has initial support from simulations which model domain walls in a de Sitter background (see Section IV.~C of \cite{Avelino:2008ve}). Though this is not the same as a domain wall dominated universe, it does suggest that a network with junctions could frustrate whenever the walls get stretched out by the Hubble flow of an accelerating universe. Taking this at face value, we will assume that the network frustrates upon domination. 
\chapter{Cosmological Perturbation Theory}
\label{ch:CosmoPert}
\chaptermark{Cosmological Perturbation Theory}

Having reviewed the basics on cosmology, inflation and domain walls, we now turn to the linearized theory of cosmological perturbations about a homogeneous and isotropic universe (see \cite{MukhanovCPT, Malik:2001rm, KodamaSasaki} for more complete reviews). After identifying the relevant degrees of freedom, we discuss the superhorizon evolution of adiabatic perturbations in a one-component universe that supports anisotropic stress gradients at large scales.

\section{Metric Perturbations}\label{ssec:metric}
We parameterize linear perturbations about a flat FRW metric as follows
\begin{align*}
ds^2 =& a^2(\tau) \left(\eta_{\mu\nu} - h_{\mu\nu}\right)dx^\mu dx^\nu \\
=& a^2(\tau) \left[ (1 + 2\varphi)d\tau^2 - 2B_i dx^i d\tau - (\delta_{ij} + h_{ij})dx^i dx^j \right]
\end{align*}
where $a(\tau)$ is the scale factor, $\tau$ is conformal time, $\eta_{\mu\nu}$ is the Minkowski metric and $h_{\mu\nu}$ represents small perturbations that depend on $x^i$ and $\tau$. In the second line, we write the perturbations in terms of the lapse function $\varphi(\tau,x^i)$, the shift vector $B_i(\tau,x^i)$ and the spatial metric $h_{ij}(\tau,x^i)$. To isolate the degrees of freedom in the metric, it is useful to split the 3-vector $B_i$ into the gradient of a scalar and a divergenceless vector
$B_i=\partial_i B + S_i$, with $\partial^i S_i = 0$. Similarly, we can split $h_{ij}$ into scalar, vector and tensor parts as follows
\begin{align*}
h_{ij} &= \frac{h}{3}\delta_{ij} + 2(\partial_i\partial_j - \frac{1}{3}\delta_{ij}\partial^2)E  + 2\partial_{(i} F_{j)} + 2h^T_{ij}\\
&= -2 \psi \delta_{ij} + 2 \partial_i\partial_j E + 2\partial_{(i} F_{j)} + 2h^T_{ij}
\end{align*}
where $h= \delta^{ij}h_{ij}$, $\partial^iF_i = \partial^i h^T_{ij}= \delta^{ij}h^T_{ij} =0$ and $\psi\equiv -\frac{1}{6} h + \frac{1}{3} \partial^2 E$ is the curvature perturbation\footnote{One can show that $a^2R_{(3)} = 4\nabla^2 \psi$, where $R_{(3)}$ is the 3-dimensional Ricci scalar associated with the induced metric on surfaces of constant $\tau$.}. Note, the 10 degrees of freedom in the metric decomposed into 4 scalar, 4 vector and 2 tensor degrees of freedom. 
\\
\indent Under a coordinate transformation $x^\mu \rightarrow x^\mu + \xi^\mu$, where $\xi^\mu =(\xi^0,\xi^i)$ and $\xi^i= \partial^i \xi + \xi_V^i$, the perturbation of a generic 4-scalar quantity $s = \bar{s} + \delta s$ transforms as (the prime denotes a derivative with respect to $\tau$)
$$\delta s\rightarrow \tilde{\delta}s = \delta s - \bar{s}'\xi^0$$
Similarly, if the background is homogeneous and isotropic, perturbations in the components of a generic 4-tensor $A^\mu_\nu = \bar{A}^\mu_\nu + \delta A^\mu_\nu$, where $\bar{A}^\mu_\nu= \text{Diag}[\bar{A}^0_0, \frac{1}{3}\delta^i_j \bar{A}^k_k]$, transform as 
\begin{align}\label{eq:TR1}
\tilde{\delta} A^0_0 =& \delta A^0_0 - \bar{A}^{0\prime}_0 \xi^0\\
\tilde{\delta} A^0_i =& \delta A^0_i + \frac{1}{3}\partial_i \xi^0 \bar{A}^{k}_k - \partial_i\xi^0 \bar{A}^0_0\\
\tilde{\delta} A^i_0 =& \delta A^i_0 - \frac{1}{3}\xi^{i\prime} \bar{A}^k_k + \xi^{i\prime} \bar{A}^{0}_0 \\\label{eq:TR2}
\tilde{\delta} A^i_j =& \delta A^i_j - \frac{1}{3}\delta^i_j \bar{A}^{k\prime}_k \xi^0
\end{align}
Thus, at linear order, both $\delta A^0_0$ and the trace of $\delta A^i_j$ ($i=j$) transform as 4-scalars, while the traceless part of $\delta A^i_j$ ($i\neq j$) remains invariant. Note, this means that $h^T_{ij}$ is gauge invariant by construction. Lastly, requiring that the line element $ds^2$ remains invariant under coordinate transformations, we can get the transformation rules for the functions $\phi$, $\psi$, $B$ and $E$ in the perturbed metric. These are \cite{Malik:2001rm}
\begin{align}\label{eq:phitransform}
\tilde{\phi} =& \phi - \CH \xi^0 - \xi^{0\prime}\\
\tilde{\psi} =& \psi + \CH \xi^0\\\label{eq:Btransform}
\tilde{B} =& B + \xi^0 - \xi'\\\label{eq:Etransform}
\tilde{E} =& E - \xi
\end{align}
where as usual $\CH=a H$ is the comoving Hubble parameter. 

\section{Matter Perturbations}
In the matter sector, because the universe is homogeneous and isotropic, the background stress-energy tensor takes the perfect fluid form $\bar{T}^\mu_\nu=(\bar{\rho}+\bar{P}) \bar{u}^\mu \bar{u}_\nu - \bar{P} \delta^\mu_\nu $, where $\bar{P}$ and $\bar{\rho}$ are the background pressure and energy density, respectively, and $\bar{u}^\mu=a^{-1}\delta^\mu_0$ is the fluid four-velocity (at the background level) for a comoving observer (note, $\bar{u}^\mu \bar{u}_\mu =1$, so $\bar{u}_\mu= a \delta^0_\mu$). In the perturbed universe we have $T^\mu_\nu = \bar{T}^\mu_\nu + \delta T^\mu_\nu$ and $u^\mu = \bar{u}^\mu + \delta u^\mu$. To first order, the deviations from the perfect fluid form appear as perturbations in the stress-energy tensor
$$\delta T^\mu_\nu = (\delta\rho + \delta P ) \bar{u}^\mu \bar{u}_\nu + (\bar{\rho} + \bar{P})(\delta u ^\mu \bar{u}_\nu + \bar{u}^\mu \delta u_\nu) - \delta P \delta^\mu_\nu - \bar{P}\Pi^\mu_\nu$$
where $\delta\rho$ is the density perturbation, $\delta P$ is the pressure perturbation, and without loss of generality the anisotropic stress $\Pi^\mu_\nu$ is chosen to be traceless and orthogonal to $u^\mu$ (so $\Pi^i_i=\Pi^0_0 = \Pi^0_i = 0$). Note, the physical anisotropic stress is actually $\bar{P}\Pi_{\mu\nu}$, but we will adopt a common abuse of language and call the dimensionless part, $\Pi_{\mu\nu}$, the ``anisotropic stress''. For the perturbed FRW metric 
to satisfy $g_{\mu\nu}u^\mu u^\nu =1$ to first order, we must set
$$u^\mu= a^{-1} (1-\varphi, v^i), \quad\quad\quad u_\mu = a(1+\varphi, -v_i - B_i)$$
where $v^i \equiv dx^i/d\tau = a \delta u^i$ is the coordinate velocity. The components of the perturbed stress-energy tensor are then
\begin{align}\label{eq:TmunuParametrization1}
\delta T^0_0&=\delta\rho\\
\delta T^i_0&= (\bar{\rho}+ \bar{P}) v^i\\
\delta T^0_i&= -(\bar{\rho}+ \bar{P}) (v_i+B_i)\\\label{eq:TmunuParametrization4}
\delta T^i_j&= -\left(\delta P \delta^i_j + \bar{P} \Pi^i_j\right)
\end{align}
Note, like the spatial metric perturbation $h_{ij}$, the spatial part of the anisotropic stress tensor can be decomposed into scalar, vector and tensor components 
$$\Pi_{ij} = \left(\partial_i\partial_j-\frac{1}{3}\delta_{ij}\partial^2\right)\Pi + \frac{1}{2}\left(\partial_i \Pi_j + \partial_j \Pi_i\right)  + \Pi^T_{ij}$$
with $\partial^i \Pi_{i}= \partial^i\Pi^T_{ij}=\delta^{ij}\Pi^T_{ij}=0$. Like $B_i$, the velocity perturbation  $v_i$ can also be decomposed as $v_i=\partial_i v + v_i^V$ into a scalar and vector part. Thus, the matter sector has 4 scalar ($\delta \rho$, $\delta P$, $v$, $\Pi$), 4 vector ($v_i^V$, $\Pi_i$) and 2 tensor ($\Pi^T_{ij}$) degrees of freedoms. Like in the metric, there are 10 degrees of freedom in the matter sector. Lastly, applying the gauge transformation rules in eq.(\ref{eq:TR1}-\ref{eq:TR2})) to $\delta T^\mu_\nu$, we can find the transformation rules for $\delta \rho$, $\delta P$, $v$ and $\Pi$. This gives
\begin{align}\label{eq:rhotransform}
\tilde{\delta} \rho =& \delta \rho - \bar{\rho}' \xi^0\\
\tilde{\delta} P =& \delta P - \bar{P}' \xi^0\\
\tilde{v} =& v + \xi'\\\label{eq:Pitransform}
\tilde{\Pi} =& \Pi
\end{align}
Thus, $\delta\rho$ and $\delta P$ transform as 4-scalars (to first order), whereas $\Pi_{ij}$ is gauge invariant (being proportional to the traceless part of $\delta T^i_j$).

\section{Gauge Invariant Variables}
As we have just seen, the perturbed universe has a total of 20 degrees of freedom. However, these are related by Einstein's equations, which reduces this number down to 10. Moreover, due to general coordinate invariance, 2 scalar and 2 vector degrees of freedom can be rotated away, leaving only 2 scalar, 2 vector and 2 tensor degrees of freedom that are physical. To ensure that perturbations are always physical, it is convenient to work with the following gauge invariant variables
\begin{align}\label{eq:GIvar1}
\Phi&\equiv\varphi + \CH\left(B-E'\right) + \left(B-E'\right)'\\\label{eq:GIvar2}
\Psi&\equiv\psi-\CH \left(B-E'\right)\\\label{eq:GIvar3}
\delta u^{(gi)}&\equiv \delta u + a(B-E')\\\label{eq:GIvar4}
\delta^{(gi)}&\equiv \delta - 3\CH\left(1+w\right)\left(B-E'\right)\\\label{eq:GIvar5}
\delta P^{(gi)}&\equiv \delta P - 3\CH \left(1+w\right)c_w^2\bar{\rho}\left(B-E'\right)
\end{align}
when dealing with scalar perturbations. The first two are known as Bardeen potentials, $\delta u = -a(v+B)$ is the scalar part of $\delta u_i$, $\delta\equiv\delta\rho/\bar{\rho}$ is the density contrast and 
\begin{equation}\label{eq:adiabaticSS}
c_w^2\equiv \frac{\bar{P}'}{\bar{\rho}'} = w  - \frac{w'}{3(1+w)\CH}
\end{equation}
is the adiabatic speed of sound (which can be derived using the Friedmann relations). Note that $c_w^2=d\bar{P}/d\bar{\rho} = w$ for a medium with an equation of state of the form $\bar{P}= w \bar{\rho}$, where $w$ is constant. Moreover, for a perfect fluid, $c_w$ turns out to be equal to $c_s$, the scalar sound speed obtained from the equations of motion for the perturbations (see Appendix \ref{app:NewtonianPert}). However, this is not automatically true in more general cases. In particular, we will see in Chapter \ref{ch:ElasticBodies} that $c_w\neq c_s$ in the case of a relativistic elastic solid.
\\
\indent It is worth mentioning that the gauge invariant variables written above coincide with the perturbations $\phi$, $\psi$, $\delta u$, $\delta$ and $\delta P$ in the Newtonian (or longitudinal) gauge where $\tilde{E}=\tilde{B}=0$. Note, one can get to this gauge by setting $\xi^0= -B + E'$ and $E =\xi$ in Eq.(\ref{eq:Btransform}) and Eq.(\ref{eq:Etransform}). Choosing to work in Newtonian gauge is often useful since any calculation done in this gauge is equivalent (but simpler) to performing the same calculation in terms of the gauge invariant variables above. Thus, the task of writing manifestly gauge invariant expressions becomes trivial once an expression is written in the Newtonian gauge.

\section{Einstein Equations}
The Einstein field equations relate metric and matter perturbations. For scalar perturbations, they can be written in gauge invariant form as follow \cite{MukhanovCPT},
\begin{align}\label{eq:EES1gi}
\nabla^2 \Psi - 3\CH (\Psi'+\CH\Phi) = & 4 \pi G a^2 \bar{\rho} \delta^{(gi)}\\\label{eq:EES2gi}
\Psi' +\CH\Phi =& 4\pi G a^2 (\bar{\rho} + \bar{P})v^{(gi)} \\\label{eq:EES3gi}
\Psi'' + \CH\left(\Phi'+2\Psi'\right) + (\CH^2 + 2 \CH')\Phi + \frac{1}{3}\nabla^2\left(\Phi - \Psi\right) =& 4 \pi G a^2 \delta P^{(gi)}\\\label{eq:EES4gi}
\Psi-\Phi =& 8 \pi G a^2 \bar{P}\Pi
\end{align}
where we have defined $v^{(gi)} \equiv a^{-1} \delta u^{(gi)}$. From the constraint $\nabla _\mu T^\mu_\nu=0$, the following energy-momentum conservation equations can then be obtained
\begin{align}\label{eq:CES1}
\delta^{(gi)\prime}=&\left(1+w\right)\left(3\Psi' + \nabla^2v^{(gi)}\right) - 3\CH\left(\frac{\delta P^{(gi)}}{\bar{\rho}}-w \delta^{(gi)}\right);\quad\quad(\nu=0)\\ \label{eq:CES2}
v^{(gi)\prime}=&-\CH\left(1-3c_w^2\right) v^{(gi)} + \frac{\delta P^{(gi)}}{1+w} +\Phi + \frac{2}{3}\frac{w}{1+w} \nabla^2\Pi;\quad\quad(\nu=i)
\end{align}
Note that Eq(\ref{eq:EES2gi}) and Eq.(\ref{eq:EES4gi}) come from trivially integrating total gradient terms. To see this, let us look at the "$i\neq j$" part of Einstein's equations:
$$(\Psi - \Phi)_{,ij} = 8 \pi  G a^2 \bar{P} \Pi_{,ij}$$
written in position space. In Fourier space (with the Liddle \& Lyth convention that we discuss shortly), it becomes
$$-k_i k_j (\Psi - \Phi) = -\left(\frac{k_ik_j}{k^2}\right)8 \pi  G a^2 \bar{P} \Pi$$
Now, we can always rotate the background coordinates so that more than one of the components of  $\vec{k}$  is non-zero, therefore we get
$$k^2(\Psi - \Phi) = 8 \pi  G a^2 \bar{P} \Pi$$
for $\vec{k}$ not equal to $\vec{0}$. The $0^{th}$ Fourier mode represents a constant offset. However, the split between background and perturbation is always chosen so that the spatial average of the perturbation vanishes and therefore equalities between gradients of perturbations imply the equality of the perturbations themselves. Going back to position space, we thus recover Eq.(\ref{eq:EES4gi}). The same argument applies to Einstein's equation $(\psi' + \CH\phi)_{,i} = 4 \pi  G a^2(\bar{\rho} + \bar{P}) v_{,i}$, from which we recover Eq.(\ref{eq:EES2gi}). 
\\
\indent Throughout the paper, we will often work in Fourier space, according to the convention used in ref.{\cite{LiddleLyth} where an extra factor of $k\equiv |\vec{k}|$ is inserted in the Fourier components of $B$, $F_i$, $v$ and $\Pi_i$, and an extra factor of $k^2$ is inserted in the Fourier components of $E$ and $\Pi$. This ensures that all perturbations in Fourier space have the same dimension and makes it easy to compare their magnitudes. In this convention, for example, we have
$$B(\tau,\vec{x})=\sum_{\vec{k}} \frac{B_{\vec{k}}(\tau)}{k}e^{i\vec{k}\cdot\vec{x}} \quad\quad\quad E(\tau,\vec{x})=\sum_{\vec{k}} \frac{E_{\vec{k}}(\tau)}{k^2}e^{i\vec{k}\cdot\vec{x}}$$
and therefore (to avoid clutter, from now on we drop the $\vec{k}$ subscript in the Fourier modes)
\begin{align*}
B_i =& i \frac{k_i}{k} B + B_i^V \\
h_{ij}=&-2 \psi \delta_{ij} - 2 \frac{k_i k_j}{k^2} E + \frac{i}{k} k_{(j} F_{i)} + 2h^T_{ij}  
\end{align*}
where now $\psi = -\frac{1}{6}h - \frac{1}{3}E$ and $ k^i B^V_i= k^i F_i = k^i h^T_{ij}=0$. Similar expressions can be found for $v_i$ and $\Pi_{ij}$. Working in Fourier space and in the Newtonian gauge (we label quantities in this gauge with a superscript $N$, with the exception of the Bardeen potentials $\Phi$ and $\Psi$), the Einstein equations for scalar perturbations become
\begin{align}\label{eq:EES1}
\CH^{-1}\Psi' + \Phi + \frac{1}{3}\left(\frac{k}{\CH}\right)^2\Psi=&-\frac{1}{2} \delta^N\\\label{eq:EES2}
\CH^{-1} \Psi' +\Phi =& -\frac{3}{2}\left(1+w\right)\frac{\CH}{k} v^N\\\label{eq:EES3}
\CH^{-2} \Psi'' + \CH^{-1}\left(\Phi'+2\Psi'\right) - 3w\Phi-\frac{1}{3}\left(\frac{k}{\CH}\right)^2\left(\Phi - \Psi\right) =& \frac{3}{2}\frac{\delta P^{N}}{\bar{\rho}}\\\label{eq:EES4}
\left(\frac{k}{\CH}\right)^2\left(\Psi-\Phi\right) =& 3 w \Pi
\end{align}
where we have used the background relations $1- \CH'/\CH^2= \frac{3}{2}(1+w)$, $4 \pi G a^2 \bar{\rho} =\frac{3}{2} \CH^2$ and $\bar{P} = w \bar{\rho}$. 
\\
\indent Note, vector perturbations decay as $a^{-2}$ in a hydrodynamical universe \cite{MukhanovBook}, therefore, they are suppressed today and we do not discuss them any further. As for tensor perturbations, there is only one Einstein equation given by
\begin{equation}\label{eq:EET}
(h^T)^{i\prime\prime}_{j} + 2\mathcal{H} (h^T)^{i\prime}_j + k^2(h^T)^i_j = 6 \CH^2 w(\Pi^T)^i_j
\end{equation}
Lastly, we note that the equations presented in this section need to be complemented by additional equations of state, defining the characteristic properties of the gravitating medium. Only after these are specified, can we solve the evolution equations for scalar and tensor modes. We discuss these in Chapter \ref{ch:ElasticBodies}.

\section{Adiabatic Perturbations}\label{sec:adiabatic}
If $\delta P$ and $\delta \rho$ satisfy $\delta P = c_w^2\delta\rho$, the perturbations in the universe are said to be adiabatic. In this case, there is only one scalar propagating degree of freedom in addition to the two tensor modes describing gravitational waves. In cosmology, this scalar adiabatic mode is typically captured by a gauge invariant variable known as the comoving curvature perturbation $\CR$. Defined as
\begin{align}
\nonumber\CR\equiv& \psi + \CH \frac{\delta u}{a}\\\label{eq:Req}
=& \psi - \CH\left(v+B\right)
\end{align}
it equals to the curvature perturbation $\psi$ in the comoving gauge $\tilde{B}=\tilde{v}=0$. This variable is useful since for a universe dominated by a single scalar field, or a perfect fluid medium with $\Pi=0$, it remains constant outside the horizon (we will see this shortly). At superhorizon scales, it also coincides with another widely used gauge invariant variable, the curvature perturbation in uniform density gauge \cite{Lyth:2004gb}
\begin{align}
\nonumber\zeta\equiv& \psi + \CH \frac{\delta \rho}{\bar{\rho}'}\\\label{eq:zetaeq}
=& \psi - \frac{\delta}{3(1+w)}
\end{align}
where to get the second line we used the background relation $\bar{\rho}' = -3\CH \bar{\rho} (1+w)$.  Note that $\zeta$ is equal to $\psi$ in the uniform density gauge where $\tilde{\delta} \rho= \tilde{B}=0$. The conservation of $\CR$ (and $\zeta$) outside the horizon, however, does not hold in a more general medium where the effects from the anisotropic stress remain important in the small $k$ limit, even if the perturbations are adiabatic. To see this, it is easiest to work in the Newtonian gauge. In this gauge, $\CR = \Psi - (\CH/k) v^N$ when written in Fourier space, and using Eq.(\ref{eq:EES2}) we obtain 
\begin{equation}\label{eq:RBardeenRelation}
\CR= \Psi +\frac{2}{3(1+ w)}\left(\CH^{-1} \Psi' + \Phi\right)
\end{equation}
Taking a derivative with respect to conformal time, we then get the general evolution equation 
\begin{align}
\nonumber \frac{3(1+w)}{2 }\frac{\CR'}{\CH} &= 3 c_w^2(\CH^{-1} \Psi' + \Phi) + \CH^{-2}\Psi'' + \CH^{-1}(\Phi' + 2\Psi') - 3w \Phi\\
\nonumber &= \left(\frac{k}{\CH}\right)^2 [-c_w^2\Psi + \frac{1}{3}\left(\Phi-\Psi\right)]+ \frac{3}{2}\left(\frac{\delta P^N}{\bar{\rho}} - c_w^2 \delta^N\right)
\end{align}
where to get the first line we used the background relation $1- \CH'/\CH^2= \frac{3}{2}(1+w)$ and Eq.(\ref{eq:adiabaticSS}), and to get the second line we used Einstein's equations Eq.(\ref{eq:EES1}) and Eq.(\ref{eq:EES3}). For adiabatic perturbations the last term vanishes, and using Eq.(\ref{eq:EES4}) the evolution equation becomes 
\begin{equation}\label{eq:Rprime}
\frac{3(1+w)}{2 }\frac{\CR'}{\CH}= -\left(\frac{k}{\CH}\right)^2 c_w^2\Psi -  w \Pi 
\end{equation}
Evidently, for a perfect fluid, $\Pi=0$ and $\CR$ is conserved at superhorizon scales. In contrast, as we will show in the next Chapter, a relativistic elastic solid is a medium where $\delta P = c_w^2\delta\rho$ but where $\Pi$ remains sizable (of the same order as $\zeta$, see Eq.(\ref{eq:EoSS})) in the $k\ll\CH$ limit. Thus, $\CR$ will not be conserved at superhorizon scales.
\chapter{Perturbations in Elastic Solids}
\label{ch:ElasticBodies}
\chaptermark{Perturbations in Elastic Solids}

The theory of elasticity studies the mechanics of a continuous medium under small external forces, which can deform the shape and volume of its body. If the body returns to its original relaxed state after being deformed, it is said to be elastic. In this section, we review the linearized theory of perturbations in relativistic elastic solids. To build intuition, we begin with the Newtonian description \cite{Landau} before moving on to the relativistic one \cite{BattyeMoss, Pearson, Sigurdson, SDM}. Lastly, as a point of comparison, we include in Appendix \ref{app:EFT} a summary of the effective field theory of broken spatial diffeomorphisms in an FRW background \cite{Lin:2015cqa}.

\section{Newtonian Description}\label{sec:ElasticBodiesNR}
Perturbations in a non-relativistic elastic solid can be described in terms of coordinates in a 3-dimensional space with an equilibrium metric $\bar{g}_{ij}$. Properties of the medium can then be unambiguously defined in this ``material'' space at some given instant of time. Small deformations of this body can be parameterized by a displacement vector $\xi^i=\xi^i(x^j)$, satisfying
\begin{equation}
\tilde{x}^i(x^j)= x^i + \xi^i(x^j)
\end{equation}
where $x^i$ denotes the relaxed position of a point in the body that has been displaced to position $\tilde{x}^i$. To first order, the distance (squared) between two infinitesimally close points after the medium is deformed, is then given by $d\tilde{s}^2=\bar{g}_{ij}(dx^i + d\xi^i)(dx^j + d\xi^j) \simeq (\bar{g}_{ik} + 2 \epsilon_{ik})dx^idx^k$, where
\begin{equation}\label{eq:StrainTensor}
\epsilon_{ik}\equiv \frac{1}{2}\left( \partial_i \xi_k + \partial_k \xi_j\right)= \partial_{(i}\xi_{j)}
\end{equation}
is the symmetric strain tensor and we have used $d\xi^i=(\partial \xi^i/\partial x^k) dx^k$. The strain tensor therefore measures the change in the relative distance between two infinitesimally close points, before and after a deformation. After the elastic body is deformed, local internal forces will restore the body to its equilibrium state. At the macroscopic level, these forces are parameterized by the divergence of the symmetric stress tensor $\sigma^{ik}$, defined by $F_i=\partial_k \sigma^{ik}$. To find the equations of motion of perturbations in the solid, one must provide a relation between the stress and strain tensors. A natural choice is to assume a linear relation of the form
\begin{equation}\label{eq:Hookeslaw}
\sigma^{ij}= E^{ijkl}\epsilon_{kl}
\end{equation}
which is the generalization of Hooke's law and where $E^{ijkl}$ is the elasticity tensor. Note, the expression for the stress tensor in Eq.(\ref{eq:Hookeslaw}) can be derived from the relation
\begin{equation}\label{eq:sigmaErelation}
\sigma^{ij}=\frac{\partial U}{\partial\epsilon_{ij}}
\end{equation}
where $U$ is the quadratic elastic potential energy function defined as
\begin{equation}\label{eq:HookeslawPot}
U=\frac{1}{2}E^{ijkl}\epsilon_{ij}\epsilon_{kl}
\end{equation}
From Eq.(\ref{eq:HookeslawPot}) and the symmetries of $\sigma_{ij}$ and $\epsilon_{kl}$, it then follows that
$E^{ijkl}$ must satisfy
\begin{equation}\label{eq:EijklSym}
E^{ijkl}=E^{(ij)(kl)}=E^{klij}
\end{equation}
The first equality in Eq.(\ref{eq:EijklSym}) reduces the number of degrees of freedom in $E^{ijkl}$ from 81 to 36, and the second equality reduces them from 36 to 21. If the body is isotropic, and we will assume this to be the case from now on, the number of degrees of freedom reduces to just 2 and $E^{ijkl}$ can be written in terms of the metric as follows \cite{Landau},
\begin{equation}\label{eq:Decomposition}
E^{ijkl} = \left(\beta - \frac{2}{3}\mu\right) \bar{g}^{ij}\bar{g}^{kl} + 2\mu \bar{g}^{i(k}\bar{g}^{l)j} 
\end{equation}
where $\beta$ is the bulk modulus and $\mu$ is the shear modulus, or rigidity. From the stress-strain relation in Eq.(\ref{eq:Hookeslaw}), with $E^{ijkl}$ written as in Eq.(\ref{eq:Decomposition}) and with $\epsilon_{kl}=\partial_{(k}\xi_{l)}$, one can then decompose $\sigma^{ij}$ as follows,
\begin{equation}\label{eq:StrainwithBulkModulus}
\sigma^{ij}= \beta \bar{g}^{ij}\partial_k\xi^k + 2\mu \left( \partial^{(i}\xi^{j)} - \frac{1}{3}\bar{g}^{ij} \partial_k\xi^k\right)
\end{equation}
The term proportional to $\mu$ vanishes when the stress tensor is purely diagonal, and the term proportional to $\beta$ vanishes when the stress tensor is purely off-diagonal. Thus, writing the stress tensor in this form suggests that, physically, $\beta$ measures the medium's resistance to a uniform compression whereas $\mu$ measures the medium's resistance to shear (volume-preserving deformations). 
\\
\indent Lastly, the equations of motion for the displacement field $\xi^i$ can be obtained from the ``$F=ma$'' formula $\bar{\rho} \ddot{\xi}^i =  \partial_j \sigma^{ij}$, where each overdot denotes a derivative with respect to time $t$, giving
\begin{equation}\label{eq:EoMNewtonian}
 \bar{\rho} \ddot{\xi}^i - \left(\beta + \frac{1}{3}\mu\right)\partial^i\partial_k\xi^k - \mu \partial_k\partial^k\xi^i=0
\end{equation}
From this equation one can then extract the scalar (longitudinal) speed of sound $c_s$ and the vector (transverse) speed of sound $c_v$
\begin{equation}\label{eq:SoS}
c_s^2=\frac{\beta}{\bar{\rho}} + \frac{4}{3}\frac{\mu}{\bar{\rho}}, \quad\quad\quad c_v^2 =\frac{\mu}{\bar{\rho}}
\end{equation}

\section{Relativistic Description} \label{sec:ElasticBodiesR}
To construct a relativistic theory of a continuous elastic medium, one must use a ``material'' manifold that is orthogonal to flow lines in the 4-dimensional space-time manifold \cite{CarterQuintana}. To this end, it is convenient to use the projected metric tensor $\gamma^{\mu\nu}=g^{\mu\nu}-u^\mu u^\nu$, where $\gamma_{\mu\nu}$ is orthogonal by construction ( $u^\mu\gamma_{\mu\nu}=0$) and therefore purely spatial. Physically, it characterizes the strain of the medium, since it can be used to find the distance between neighboring ``particles'' that are locally at rest. It can also be used to construct other orthogonal tensors, such as the pressure tensor $P^{\mu\nu}$ or the elasticity tensors $E^{\mu\nu\alpha\beta}$.
\\
\indent  In analogy with the Newtonian description, the defining characteristic of the relativistic elastic solid is that the pressure tensor is a function of the strain. Assuming the medium to be isotropic, the material tensors can then be decomposed as
\begin{align}
P^{\mu\nu} =& \bar{P} \gamma^{\mu\nu}\\\label{eq:DecompositionRel}
E^{\mu\nu\alpha\beta} =& \left(\beta-\bar{P}- \frac{2}{3}\mu \right) \gamma^{\mu\nu}\gamma^{\alpha\beta} + 2\left(\mu+\bar{P}\right) \gamma^{\mu(\alpha}\gamma^{\beta)\nu} 
\end{align}
since $\gamma^{\mu\nu}$ is the only isotropic tensor available which is orthogonal. Note, $u_\mu P^{\mu\nu}=u_\mu E^{\alpha\beta \mu\nu}=0$ and the material tensors are also orthogonal, as they must be in order to define intrinsic properties of the medium. In fact, under these assumptions, it was shown in \cite{BattyeMoss} (and reviewed in \cite{Pearson}) that specifying these two material tensors is all one needs to calculate the variation of the energy-momentum tensor (which sources the gravitational field equations), and therefore, to describe the behavior of linear perturbations in a relativistic elastic solid. Working in the synchronous gauge ($\delta g_{\mu\nu} = -a^2(\tau) h_{\mu\nu}$, $h_{00} =h_{0i}=0$) the perturbed energy-momentum tensor of an isotropic relativistic elastic solid takes the following form \cite{Pearson}
\begin{align}\label{eq:TzerozeroNotOrthogonal}
\delta T^0_0=&-(\bar{\rho}+\bar{P})\left(\partial_k\xi^k+\frac{h}{2}\right) -\left[\bar{\rho}' + 3\CH\left(\bar{\rho}+\bar{P}\right)\right]\xi^0\\
\delta T^i_0=&(\bar{\rho}+\bar{P})\xi^{i\prime}\\\label{eq:TijNotOrthogonal}
\delta T^i_j=& \beta \left(\partial_k \xi^k + \frac{h}{2}\right)\delta^i_j + 2\mu \left( \partial_{(j}\xi^{i)} + \frac{h^i_j}{2}  - \frac{1}{3}\left(\partial_k \xi^k + \frac{h}{2}\right)\delta^i_j\right) + \left(\bar{P}' + 3\beta\CH \right)\xi^0\delta^i_j
\end{align}
where $\xi^\mu=(\xi^0,\xi^i)$ represents small deformations relative to the fixed background space. By demanding that the solid only fluctuates in space (i.e. after imposing time diffemorphism invariance) it follows that the terms proportional to $\xi^0$ in Eq.(\ref{eq:TzerozeroNotOrthogonal}) and Eq.(\ref{eq:TijNotOrthogonal}) must vanish. This constraint fixes the evolution of $\bar{\rho}$ and $\bar{P}$, which now must obey the following equations
\begin{align}\label{eq:rhoconstraint}
\bar{\rho}' + 3\CH\left(\bar{\rho}+\bar{P}\right)=&0\\\label{eq:Pconstraint}
\bar{P}' + 3\beta\CH =0
\end{align}
The first equation is the usual background continuity equation and the second equation allows us to solve for the relativistic bulk modulus, giving
\begin{align}\label{eq:beta}
\beta = \left(\bar{\rho}+\bar{P}\right) \frac{d \bar{P}}{d\bar{\rho}}
\end{align}
\\
\indent The perturbed momentum conservation equation $\gamma^\nu_\alpha\delta\left(\nabla_\mu T^\mu_\nu\right)=0$ yields the following equation of motion for the displacement field \cite{Pearson}
\begin{equation}\label{eq:EoMRel}
 \left(\bar{\rho}+\bar{P}\right)\left[ \xi^{\prime\prime i} +\CH \xi'^i\right] - 3 \beta \CH \xi'^i - \left(\beta + \frac{1}{3}\mu\right)\partial^i\partial_k\xi^k - \mu \partial_k\partial^k\xi^i= \frac{1}{2}\left(\beta -\frac{2}{3}\mu\right) \partial^i h + \mu\partial^k h^i_k
 \end{equation}
Compared to the Newtonian analog in Eq.(\ref{eq:EoMNewtonian}), the relativistic equation of motion has a source term due to the spatial metric perturbations and a friction term due to the Hubble flow. It also contains terms proportional to the scalar pressure $\bar{P}$, accounting for the fact that the background pressure can now be of the same order as the background energy density $\bar{\rho}$. Note that in this (synchronous) gauge, we recover all of the non-relativistic expressions obtained in Section \ref{sec:ElasticBodiesNR}, simply by taking the limit where $h^i_j\ll\partial_{j}\xi^{i}$, $\bar{P}\ll\bar{\rho}$, and $\CH\rightarrow 0$. To find expressions for the scalar and vector sound speeds, one can decompose Eq.(\ref{eq:EoMRel}) into its scalar and vector components and get \cite{BattyeMoss, Carter}
\begin{equation}\label{eq:SoSRel}
c^2_s= \frac{d \bar{P}}{d\bar{\rho}} + \frac{4}{3}c_v^2, \quad\quad\quad\quad c^2_v = \frac{\mu}{\bar{\rho}+\bar{P}}
\end{equation} 
We see that, even if $d\bar{P}/d\bar{\rho}<0$, $c_s^2$ can be positive if the rigidity parameter $\mu$ is large enough. Note, this is not the case for perfect fluids; instabilities arise at small scales if $w<0$, since we instead have $c_s^2=c_w^2=w$.
\\
\indent Expressions for $\delta \rho$, $v^i$, $\delta P$ and $\Pi^i_j$ in terms of $\xi^i$ can be found by comparing the perturbed energy-momentum tensor in Eq.(\ref{eq:TzerozeroNotOrthogonal}-\ref{eq:TijNotOrthogonal}), subject to the constraints in Eq.(\ref{eq:rhoconstraint}-\ref{eq:Pconstraint}), with the decomposition written in Eq.(\ref{eq:TmunuParametrization1}-\ref{eq:TmunuParametrization4}). This yields
\begin{align}\label{eq:EoS1}
\delta\rho=&-(\bar{\rho}+\bar{P})\left(\partial_k\xi^k + \frac{h}{2}\right)\\\label{eq:EoS2}
v^i=&\xi^{i\prime}\\\label{eq:EoS3}
\delta P =&\frac{d\bar{P}}{d\bar{\rho}}\delta\rho\\ \label{eq:EoS4}
\Pi^i_j =& -\frac{\mu}{\bar{P}} \left( 2\partial_{(j}\xi^{i)} + h^i_j  - \frac{2}{3}\delta^i_j\left(\partial_k \xi^k + \frac{h}{2}\right)\right)
\end{align}
We see from Eq.(\ref{eq:EoS3}) that, like for the perfect fluid, the scalar perturbations in the solid are adiabatic. Unlike a perfect fluid, however, $d\bar{P}/d\bar{\rho}\neq c_s^2$, where $c_s^2$ is the scalar sound speed derived in Eq.(\ref{eq:SoSRel}). This is because $c_s^2$ now has an extra contribution coming from the rigidity of the solid.
\\
\indent The four equations above can be decomposed into scalar and tensor components. After some simple manipulations, the tensor and scalar part of the anisotropic stress can be written as \cite{Sigurdson}
\begin{equation}\label{eq:EoST}
(\Pi^T)^i_j= -\frac{\tilde{\mu}}{w}(h^T)^i_j
\end{equation}
\begin{equation}\label{eq:EoSS}
\nabla^2 \Pi = -\frac{6\tilde{\mu}}{w}\zeta
\end{equation}
which are manifestly gauge invariant (since both $h^T_{ij}$ and $\zeta$ are gauge invariant) and where we have defined $\tilde{\mu}\equiv \mu/\rho$. Eq.(\ref{eq:EoST}) and Eq.(\ref{eq:EoSS}) are the \emph{characteristic equations of state encoding the elastic properties of the medium}. We see that the amplitude of the Fourier modes for $(\Pi^T)^i_j$ and $\Pi$ are always of the same order as the amplitude of the gravitational waves and curvature perturbations, respectively, that propagate through the elastic medium. This is true regardless of the value for $k/\CH$, so the perturbations in the medium will evolve at all scales, including those where $k\ll \CH$. We will see in the next Chapter that the anisotropic stress at large scales effectively acts as a mass for perturbations; perturbations will decay at superhorizon scales.
\chapter{Superhorizon Evolution of Perturbations}
\label{ch:EvoPert}
\chaptermark{Superhorizon Evolution of Perturbations}
In this chapter, we derive the superhorizon evolution equations for tensor and scalar perturbations in a universe that behaves like a relativistic elastic solid at large scales. We then derive the matching conditions that need to be imposed on these perturbations when the universe transitions between fluid-like and solid-like phases.

\section{Solid-like Universe}\label{sec:soliduniverse}
The evolution of the tensor mode can be easily found since there is only one Einstein equation, Eq.(\ref{eq:EET}). From Eq.(\ref{eq:EoST}), we then have
\begin{equation}\label{eq:EoMgamma2}
(h^T)^{i\prime\prime}_{j} + 2\CH (h^T)^{i\prime}_j + \left(k^2 +  4 c_v^2\epsilon \CH^2\right) (h^T)^i_j =0
\end{equation}
where $\epsilon = 3/2(1+w)$ and we used $\tilde{\mu} = 2/3c_v^2 \epsilon$. To find the evolution equation for the scalar mode, we make use of the following relations between gauge invariant variables
\begin{align}\label{eq:Eprime00}
v^{(gi)} &= \CH^{-1}\left( \CR - \Psi\right)\\\label{eq:Eprime01}
\delta^{(gi)} &= 2\epsilon \left(\Psi - \zeta\right)
\end{align}
which follow trivially from their definitions. Plugging Eq.(\ref{eq:Eprime00}-\ref{eq:Eprime01}) into Einstein equations Eq.(\ref{eq:EES1gi}-\ref{eq:EES2gi}), and using and Eq.(\ref{eq:RBardeenRelation}), we get that (going to Fourier space) 
\begin{equation}\label{eq:Eprime1}
\left(\frac{k}{\CH}\right)^2 \Psi = 3\epsilon\left(\zeta -\CR\right)
\end{equation}
Note that in any gauge with $\psi=0$, the Bardeen variable $\Psi\equiv \psi - \CH(B-E')$ becomes proportional to the ``geometric shear'', $\sigma_s\equiv E' -B$, which is a measure of the anisotropic component of the universe's expansion. Note as well that $\zeta/\CH-\CR/\CH$ is equal to the time diffeomorphism needed to go from the uniform density gauge to the comoving gauge. Thus, Eq.(\ref{eq:Eprime1}) gives the intuitive result that constant-$\rho$ time-slices are orthogonal to the 4-velocity of the medium only if there is no anisotropic component present at large scales. This is certainly the case when the universe is dominated by a perfect fluid or during single-field slow-roll inflation, where the anisotropic stress vanishes and $\CR\simeq \zeta$ at large scales. As we have seen, however, the anisotropic stress in a solid-like universe remains important at large scales and we therefore expect that $\zeta \neq \CR$.
\\
\indent Plugging Eq.(\ref{eq:Eprime00}-\ref{eq:Eprime01}) into the perturbed continuity relation, Eq.(\ref{eq:CES1}), we find that
\begin{equation}\label{eq:Eprime2}
\left(\frac{k}{\CH}\right)^2 \Psi= -3\frac{\zeta'}{\CH} + \left(\frac{k}{\CH}\right)^2 \CR
\end{equation}
which, when combined with Eq.(\ref{eq:Eprime1}), gives a the following useful result
\begin{equation}\label{eq:Rzetarelation}
\CR= \frac{1}{\epsilon \CH}\frac{\zeta' + \epsilon\CH \zeta}{1+ k^2/3\epsilon\CH^2}
\end{equation}
Note that in the $k\ll\CH$ limit we have $\CR -\zeta \simeq \zeta'/\epsilon\CH$ and we get from Eq.(\ref{eq:Eprime1}) that $\zeta'\simeq\CR'\simeq0$ in the absence of an anisotropic component at superhorizon scales (the left-hand-side of Eq.(\ref{eq:Eprime1}) vanishes in this case). Note also that we have not yet made use of the equations of state of the solid, so the above results apply to adiabatic perturbations in any medium that can support an anisotropic component at large scales.
\\
\indent We now derive the evolution equation for the scalar mode $\zeta$ in a solid-like universe. Using Eq.(\ref{eq:Rprime}), Eq.(\ref{eq:Eprime1}), the time derivative of Eq.(\ref{eq:Rzetarelation}), the elastic equation of state Eq.(\ref{eq:EoSS}), and taking the $k\ll \CH$ limit, we obtain 
\begin{equation}\label{eq:EoMzeta}
\zeta'' + 2\CH \zeta'+ 4 c_v^2 \epsilon  \mathcal{H}^2 \zeta = 0, \quad\quad\quad (k\ll \CH)
\end{equation}
We have found that, at large scales, $\zeta$ and $(h^T)^i_j$ evolve in the same way during the solid-like period. Note, unlike adiabatic perturbations in standard cosmologies (with $\Pi_{ij}=0$), the scalar and tensor modes \textit{decay} at superhorizon scales due to the extra time-dependent mass term proportional to $c_v^2$, and thus the rigidity $\mu$. Physically, this superhorizon suppression is due to \textit{local} anisotropic pressure gradients in the elastic universe, so there are no acausal effects (as a naive interpretation of the word ``superhorizon'' might suggest).
\\
\indent An evolution equation for $\CR$ can also be obtained. Combining the momentum conservation relation in Eq.(\ref{eq:CES2}) and the scalar equation of state for the solid, Eq.(\ref{eq:EoSS}), we get
\begin{equation}\label{eq:zetaRrelation}
\zeta = \frac{c_w^2}{c_s^2}\CR - \frac{1}{3 c_s^2 \mathcal{H}}\CR'
\end{equation}
where $c_s^2$ is given in Eq.(\ref{eq:SoSRel}) and $c_w^2=dP/d\rho$ is given in Eq(\ref{eq:adiabaticSS}). Combining Eq.(\ref{eq:Rzetarelation}) with Eq.(\ref{eq:zetaRrelation}), we then obtain
\begin{equation}\label{eq:EoMR0}
\CR''+\left(2+\eta_{\epsilon}- \eta_s\right)\CH \CR' 
+ c_s^2k^2\CR 
+ 4c_v^2\CH^2\left( \epsilon + \eta_v - \eta_s \right) \CR = 0
\end{equation}
where $\eta_\epsilon \equiv \epsilon'/(\epsilon\CH)$, $\eta_s \equiv (c_s^2)'/(c_s^2 \CH)$ and $\eta_v \equiv (c_v^2)'/(c_v^2 \CH)$. Note, this equation was also derived in \cite{Sigurdson}. 

\section{Fluid-to-Solid Transition}\label{sec:matching}
We now derive the matching conditions that need to be imposed on cosmological perturbations transitioning between fluid-like and solid-like universes. We will assume that the transition is much shorter than a Hubble time (we will justify this assumption later in Chapter \ref{ch:Model}). Therefore, the matching surface can be defined on a constant-$\tau$ hypersurface $\Sigma$, where the conformal time is $\tau=\tau_{-}$ on one side and $\tau=\tau_{+}$ on the other. We also pick coordinates such that the local energy density $\rho(\tau,x^i)$ is constant on the space-like surface $\Sigma$.
\\
\indent In an FRW universe, to zeroth order, the matching surface $\rho = \bar{\rho}$ already coincides with a constant-$\tau$ hypersurface. In the perturbed universe, however, constant-$\rho$ and constant-$\tau$ surfaces do not coincide in general. For them to coincide, we must choose the following coordinate system
\begin{equation}\label{eq:gauge1}
\tau \rightarrow \tilde{\tau} = \tau + \xi^0=const.
\end{equation}
\begin{equation}\label{eq:gauge2}
\delta \rho  \rightarrow \tilde{\delta}\rho =\delta\rho -\bar{\rho}' \xi^0\ = 0
\end{equation}
where the temporal gauge is fixed to $\xi^0=\delta\rho/\bar{\rho}'$ and the spatial gauge $\xi$ remains unspecified. Matching conditions for the cosmological perturbations can be obtained after imposing the continuity of the induced 3-metric $\gamma_{\mu\nu}$ on the hypersurface $\Sigma$, and the continuity of the extrinsic curvature $K_{\mu\nu}= \gamma^\lambda_\mu \nabla_\lambda n_\nu$ on $\Sigma$, where $n^\mu$ is the normal vector to the hypersurface \cite{Israel}. Here, $K_{\mu \nu}$ is a 3-dimensional spatial tensor since $n^\mu$ is normal to $\Sigma$ \cite{Wald}. In general, however, the extrinsic curvature is given by $K_{\mu\nu} = \gamma_\mu^\alpha \gamma_\nu^\beta \nabla_{(\alpha}u_{\beta)}$ or $K_{\mu\nu} = \frac{1}{2}\mathcal{L}_u g_{\mu \nu}$, where $\mathcal{L}_u$ is the Lie derivative along $u^\mu$. The continuity of $\gamma_{\mu\nu}$ and $K_{\mu\nu}$ are called the ``Israel junction conditions'' and they follow from the natural requirement that the intrinsic geometry on $\Sigma$ be the same as we approach the matching surface from either side.
\\
\indent In these coordinates (denoted by ``tilde''), the scalar part of the linearized FRW metric given in Section \ref{ssec:metric} splits into the following trace and traceless components 
\begin{equation}\label{eq:gammaij}
\tilde{\gamma}_{ij}=-a^2\left\{\left(1+ \frac{\tilde{h}}{3}\right)\delta_{ij} + 2\left(\partial_i\partial_j - \frac{1}{3}\delta_{ij}\partial^2\right)\tilde{E} \right\}
\end{equation}
Similarly, using the linearized expression for the normal vector $n^\mu= a^{-1}\left[\left(1 -\varphi\right), \partial^iB\right]$ and the connection coefficient $\Gamma_{j 0}^i=\CH \delta^i_j - \psi'\delta^i_j + \partial_j\partial^i E'$, the scalar part of the extrinsic curvature in these coordinates reads
\begin{align}
\nonumber \tilde{K}^i_j =& \partial_j n^i - \Gamma_{j 0}^i n^0\\\label{eq:Kij}
 =& -\frac{1}{a}\left\{ \left[\CH\left(1-\tilde{\varphi}\right) - \tilde{\psi}' + \frac{1}{3}\partial^2 \tilde{\sigma}_s\right]\delta^i_j + \left[\partial^i\partial_j - \frac{1}{3} \delta^i_j \partial^2\right]\tilde{\sigma}_s\right\}
\end{align}
Like the induced metric, $\tilde{K}^i_j$ can be split into its trace and traceless part. Physically, the trace part gives the isotropic component of the local expansion, whereas the traceless part gives the anisotropic component $\tilde{\sigma}_s\equiv \tilde{E}' -\tilde{B}$, which is the geometric shear.
\\
\indent To zeroth order, the first junction condition $[\tilde{\gamma}_{ij}]^{\pm}=0$ immediately implies the continuity of the scale factor $a$, or equivalently, that $\left[a\right]^{\pm}\equiv a(\tau_{+}) - a(\tau_{-}) = 0$. Similarly, the second junction condition $\left[\tilde{K}_{ij}\right]^{\pm}=0$, implies that $\left[\CH\right]^{\pm}=0$ and therefore that $a'$ is also continuous on $\Sigma$. To first order in the scalar perturbations, we get
\begin{align}\label{eq:cond1}
\left[\tilde{\psi}\right]^{\pm}=&0\\\label{eq:cond2}
\left[\tilde{\sigma}_s\right]^{\pm}=&0 \\\label{eq:cond3}
\left[\CH\left(1 - \tilde{\varphi}\right) - \tilde{\psi}'\right]^{\pm}=&0
\end{align}
Eq.(\ref{eq:cond1}) follows from the continuity of $\tilde{\gamma}_{ij}$, as given in Eq.(\ref{eq:gammaij}), using the fact that the trace and traceless part of the induced metric, $\tilde{h}$ and $\tilde{E}$, are separately continuous and that the intrinsic curvature perturbation is $\tilde{\psi} = -\frac{1}{6} \tilde{h} + \frac{1}{3}\partial^2 \tilde{E}$. Similarly, Eq.(\ref{eq:cond2}) and Eq.(\ref{eq:cond3}) follow directly from the continuity of $\tilde{K}^i_j$, as given in Eq.(\ref{eq:Kij}). Note, the above results are valid in the temporal gauge given by Eq.(\ref{eq:gauge1}) and Eq.(\ref{eq:gauge2}). Using Eq.(\ref{eq:phitransform}-\ref{eq:Etransform}), however, we can rewrite the matching conditions in arbitrary coordinates \cite{Deruelle:1995kd}
\begin{align}\label{eq:MC1}
\left[\psi + \CH \frac{\delta \rho}{\bar{\rho}'}\right]^{\pm}=&0\\ \label{eq:MC2}
\left[B- E' + \frac{\delta \rho}{\bar{\rho}'}\right]^{\pm}=&0\\\label{eq:MC3}
\left[\CH\phi + \psi' + \left(\CH'-\CH^2\right)\frac{\delta \rho}{\bar{\rho}'}\right]^{\pm}=&0
\end{align}
From Eq.(\ref{eq:zetaeq}) we can write Eq.(\ref{eq:MC1}) as $\left[\zeta\right]^{\pm}=0$, so $\zeta$ is continuous. Note that in a gauge where $\psi=0$, the continuity of $\zeta$ implies that $\delta_{+}(1+w_{+})^{-1} = \delta_{-}(1+w_{-})^{-1}$ and thus that no entropy perturbations are generated at the transition between the fluid and solid phases (see \cite{Sigurdson}). Similarly, we can also write Eq.(\ref{eq:MC2}) as $\left[\zeta - \Psi\right]^{\pm}=0$, where $\Psi$ is the Bardeen potential written in Eq.(\ref{eq:GIvar2}), and therefore $\Psi$ is also continuous. This must actually be the case. Indeed, if $\Psi$ underwent a sharp jump then the $\Psi''$ term in Eq.(\ref{eq:EES3}) would imply the presence of a derivative of a delta function that cannot be removed. Lastly, we can use Eq.(\ref{eq:GIvar1}), Eq.(\ref{eq:GIvar2}), Eq.(\ref{eq:zetaeq}) and Eq.(\ref{eq:RBardeenRelation}) to rewrite the last matching condition in Eq.(\ref{eq:MC3}) as
\begin{equation}\label{eq:MC4}
\left[\frac{3(1+w)}{2}\left(\CR - \zeta\right)\right]^{\pm}=0
\end{equation}
From Eq.(\ref{eq:Eprime1}) in the $k\ll \CH$ limit, we thus have
\begin{equation}\label{eq:MCzeta}
\left[\zeta'\right]^{\pm}\simeq 0
\end{equation}
and $\zeta'$ is continuous at large scales. We can also obtain two matching conditions involving $\CR$ and $\CR'$. The first condition follows from Eq.(\ref{eq:zetaRrelation}) and the continuity of $\zeta$. This gives
\begin{equation}\label{eq:MCR1}
\left[\frac{w}{c_s^2}\CR - \frac{\CR'}{3c_s^2\CH}\right]^{\pm} = 0
\end{equation}
The second matching condition follows from Eq.(\ref{eq:MC4}) and Eq.(\ref{eq:zetaRrelation}), which when combined gives
\begin{equation}\label{eq:MCR2}
\left[ \left(\frac{1+w}{ c_s^2}\right)\left(\frac{\CR'}{\CH} + 4c_v^2\CR\right)  \right]^{\pm} =0
\end{equation}
As expected, for a medium with no rigidity ($\mu=c_v^2=0$ and $c_s^2=w$), we recover the matching conditions for the transition between two fluid-like universes, as given in ref.\cite{Namjoo:2012xs}.
\\
\indent Matching conditions for the tensor perturbations are easier to obtain, since we just have $\tilde{\gamma}_{ij}= -a^{2} h^T_{ij}$ and $K^i_j= -\frac{1}{2} a^{-1} h^{Ti\prime}_j$. Therefore, the matching conditions are simply
\begin{equation}
\left[h^T_{ij}\right]^{\pm}=\left[h^{T\prime}_{ij}\right]^{\pm}=0
\end{equation}
and since $h^T_{ij}$ is already gauge invariant, these conditions are valid in any coordinate frame. 
\\
\indent In this chapter, we have shown that $h$ and $\zeta$ obey the same evolution equations as well as the same trivial matching conditions. For this reason, we will use these variables to track the superhorizon evolution of scalar and tensor modes. Note, we could use $\CR$ instead of $\zeta$ to track the scalar mode, but this would require more work since the equations of motion and matching conditions are more complicated in this case. Ultimately, however, the final result in the late-time, large-scale limits would be the same. 
\chapter{Observables in a Cosmology with a Domain Wall Era}
\label{ch:DWCosmo}
\chaptermark{Power Spectra in Cosmology with a Domain Wall Era}

We now turn to the cosmology shown in Fig.(\ref{fig:CosmicHistory}), where a domain wall network produced shortly after inflation, dominates the energy density for a few e-folds before decaying and reheating the universe. 
We will use the results from the linear theory of relativistic elastic solids, obtained in Chapter \ref{ch:EvoPert}, to track the evolution of perturbations during domain wall dominance. We will assume that the standard radiation era follows after the network decays, and that the transition to and from domain wall dominance occurs instantaneously. In Section \ref{sec:Model}, we will justify these assumptions in the context of a specific model. Lastly, We briefly discuss non-gaussianities. 

\section{Late-time Power Spectra} 
Let $\tau_*$, $\tau_{1}$ and $\tau_{2}$ denote the (conformal) time when the mode $k_*$ exited the horizon during inflation, the time when the domain wall network starts to dominate, and the time when the network decays, respectively. Before domain wall dominance, the perturbations are frozen at superhorizon scales (since there is no anisotropic stress; see Section \ref{sec:soliduniverse}) and thus $\zeta(\tau_1)\simeq\zeta_*$, where $\zeta_*\equiv \zeta(\tau_*)$ is the perturbation at horizon exit. To find $\zeta$ during domain wall dominance, we must solve the evolution equation
\begin{equation}\label{eq:EoMzeta2}
\zeta'' + \frac{4}{1+3w}\frac{\zeta'}{\tau} + \frac{24 \tilde{\mu}}{(1+3w)^2} \frac{\zeta}{\tau^2} = 0
\end{equation}
where we made use of Eq(\ref{eq:SoSRel}), $\epsilon = 3(1+w)/2$, $c_v^2 =3/2\tilde{\mu}/\epsilon$, and the background relation $\CH = 2(1+3 w)^{-1}\tau^{-1}$ to recast Eq.(\ref{eq:EoMzeta}). Solving  this equation and using the continuity of $\zeta$ and $\zeta'$ at $\tau = \tau_1$, we obtain
\begin{equation}\label{eq:zetasols1}
\zeta(\tau_{1}\leq\tau<\tau_2) =  \zeta_{*} \left(\frac{\tau}{\tau_1}\right)^{\beta}\left[\cos\left(\nu\log\frac{\tau}{\tau_1}\right) - \frac{\beta}{\nu}\sin\left(\nu\log\frac{\tau}{\tau_1}\right)\right]
\end{equation}
where 
\begin{equation}
\beta = \frac{3}{2}\left(\frac{w -1}{1 + 3w}\right),\qquad\qquad
\nu= -\frac{\sqrt{96\tilde{\mu} - 9(w-1)^2}}{2(1+3w)}
\end{equation}
\indent At $\tau=\tau_2$, the walls decay and we assume the universe becomes radiation dominated. Now Eq.(\ref{eq:EoMzeta2}) with $w=1/3$ and $\tilde{\mu}=0$ has the simple solution
\begin{equation}
\zeta(\tau>\tau_2)=\zeta(\tau_2)+{D \over \tau}.
\end{equation}
for some integration constant $D$. Imposing the continuity of $\zeta$ and $\zeta'$ at $\tau=\tau_2$, and discarding the decaying piece, one gets
\begin{equation}\label{eq:latetimesfluctuation}
\zeta(\tau\gg\tau_2) \simeq \zeta_{*} \left(\frac{\tau_2}{\tau_1}\right)^{\beta}\left[\cos\left(\nu\log\frac{\tau_2}{\tau_1}\right)  - \frac{\beta+ \beta^2+\nu^2}{\nu}\sin\left(\nu\log\frac{\tau_2}{\tau_1}\right) \right]
\end{equation}
As expected, when the solid-like period disappears ($\tau_2\rightarrow\tau_1$), or  when the anisotropic stress vanishes ($\tilde{\mu}\rightarrow0$, so $\beta\rightarrow i\nu$), we recover the standard result where $\zeta(\tau\gg\tau_2)\simeq \zeta_*$ and the perturbations remain frozen outside the horizon.
\\
\indent As mentioned in Section \ref{ch:Background}, for the solid to describe a frustrated domain wall network, we must set $w=-2/3$ and \cite{Rigidity} $\tilde{\mu}=4/15$. In this case, $(\tau_2/\tau_1)=e^{-\N_{dw}/2}$, $\beta= 5/2$ and $\nu = (1/2)\sqrt{3/5}$. Therefore,
\begin{equation}
\mathcal{P}_{\zeta,0}(k_*)\simeq \frac{k_*^3}{2\pi^2}|\zeta(\tau\gg\tau_2)|^2 \simeq \mathcal{T} \times \mathcal{P}_{\zeta}(k_*)
\end{equation}
where $\mathcal{P}_{\zeta,0}(k_*)\sim 10^{-9}$ is the power spectrum today (at the pivot scale $k_*\simeq0.05$ Mpc$^{-1}$), $\mathcal{P}_{\zeta}= k_*^3|\zeta_*|^2/(2\pi^2)$ is the primordial power spectrum at horizon exit and
\begin{equation}\label{eq:T}
\mathcal{T}(\N_{dw})= e^{-\frac{5}{2} \N_{dw}} \left[\cos\left(\frac{1}{4}\sqrt{\frac{3}{5}}\N_{dw}\right) + \frac{89}{\sqrt{15}}\sin\left(\frac{1}{4}\sqrt{\frac{3}{5}}\N_{dw}\right)\right]^2 
\end{equation}
is the \textit{transfer function} encoding the effects from the domain wall period (see Fig.(\ref{fig:LTO})). Note, the sharp modulation coming from the oscillatory term in $\mathcal{T}$ is an artifact of the instantaneous transition approximation; 
we expect this effect to smooth out as the transition between epochs becomes longer.
\begin{figure}[t]
\center{
\includegraphics[scale=0.3]{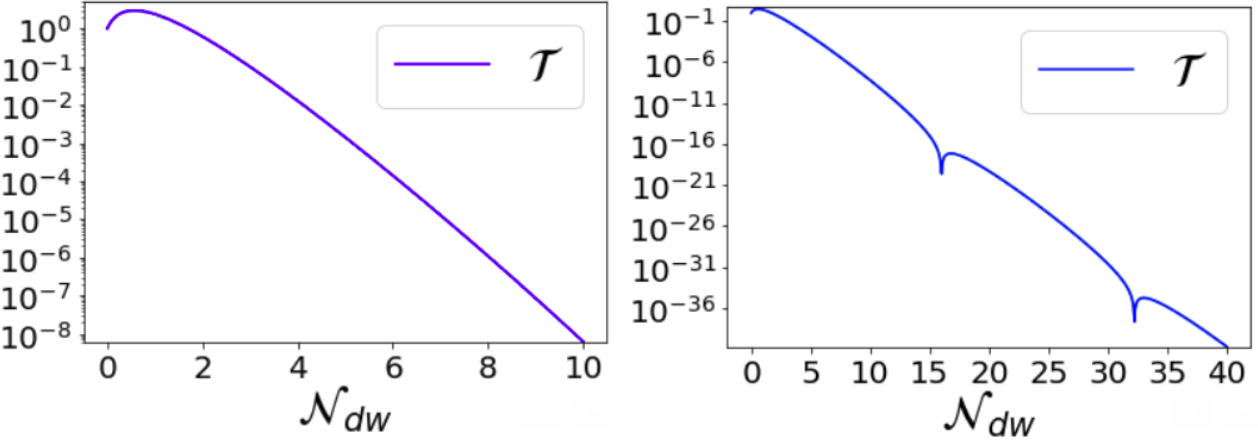}
\caption{Semilog plot of the transfer function $\mathcal{T}(\N_{dw})$ in Eq.(\ref{eq:T}) ($w=-2/3$ and $\tilde{\mu}=4/15$).
\label{fig:LTO}}}
\end{figure}
\\
\indent Since $\zeta$ and $h$ evolve in the same way outside the horizon, the above result also applies to tensor fluctuations. Using the standard predictions from single-field slow-roll inflation that we derived in \ref{eq:primordialscalarinf} and \ref{eq:primordialtensorinf}), we obtain
\begin{equation}\label{eq:Normalization}
\mathcal{P}_{\zeta}(k_*) = \frac{H_i^2}{8 \pi^2 M_p^2 \epsilon_*} = \frac{\mathcal{P}_{\zeta,0}(k_*)}{\mathcal{T}}
\end{equation} 
\begin{equation}\label{eq:Normalizationh}
\mathcal{P}_{h}(k_*) = \frac{2H_i^2}{\pi^2M_p^2} = \frac{\mathcal{P}_{h,0}(k_*)}{\mathcal{T}} 
\end{equation} 
where $M_p$ is the reduced Planck mass and $H_i$ is Hubble during inflation. Note, even though the tensor-to-scalar ratio $r_*\equiv\mathcal{P}_{h}(k_*)/\mathcal{P}_{\zeta}(k_*)$ is independent of $\mathcal{T}$, the observational bound \cite{Planck} $r_*< 0.07$ combined with the constraint $\mathcal{P}_{\zeta}\sim 10^{-9}/\mathcal{T}$, places the following bound on the Hubble scale:
\begin{equation}\label{eq:energyscalebound}
H_i < \frac{10^{14} \text{ GeV}}{\mathcal{T}^{1/2}}
\end{equation} 
This is the main result in this thesis. We have found that the main effect of the domain wall network is to exponentially suppress the power spectra at superhorizon scales and that, interestingly, this suppression relaxes the current bounds on $H_i$ by a factor of $\mathcal{T}^{-1/2}$. The only bound on $H_i$ would come from considerations of eternal inflation.  Saturating this bound, the inflationary scale can take values close the Planck scale $M_p$.
\\
\indent As a simple example, let us now consider the chaotic inflation model introduced in Section \ref{sec:InfCosmo}, where the inflaton potential takes the form $V= \frac{1}{2}m^2\phi^2$, $m$ being the inflaton mass. It is straightforward to show that in this case we have
\begin{equation}\label{eq:Pchaotic}
\mathcal{P}_{\zeta}(k_*) \simeq \frac{m^2}{M_p^2} \frac{\N_*^2}{6\pi^2} \sim \frac{10^{-9}}{\mathcal{T}}
\end{equation}
\begin{equation}\label{eq:nschaotic}
n_s \simeq 1-\frac{2}{\N_*} \sim 0.96
\end{equation}
where $\N_*$ is the number of e-folds elapsed between the horizon exit of the pivot scale and the end of inflation and $n_s$ is the spectral index. We see that to satisfy Eq.(\ref{eq:nschaotic}) we must have\footnote{Note, if $\N_{dw}$ is significantly larger than 10, then the number of e-folds needed to solve the horizon problem can be considerably less than the usual 50-60 e-folds. For our purposes, however, we can ignore this effect since $\N_{dw}<10$.} $\N_*\sim 50$ and to satisfy Eq.(\ref{eq:Pchaotic}) we must have $m\sim 7\times 10^{-6}M_p \mathcal{T}^{-1/2}$. For $\N_{dw} = 7$, this gives $m\sim 0.02 M_p$, which is over three orders of magnitude larger than the value of $m$ obtained in the cosmology without the domain wall era.

\section{Late-time Non-Gaussianities}
Let us finish this chapter with a brief description of non-Gaussianities in this cosmology. For single-field slow-roll inflation, it was shown in \cite{Maldacena:2002vr} that $\langle \zeta^3\rangle \sim f_{\text{NL}} \mathcal{P}^2_{\zeta}(k_*)$ at horizon exit,
where $f_{\text{NL}}\sim\mathcal{O}(\epsilon, \eta)$ is a slow-roll suppressed constant that measures the size of the bispectrum in the equilateral momentum configuration (note, since $f_{\text{NL}}$ depends only on the overall momentum scale, it is constant if the primordial power spectrum is scale invariant). The primordial non-Gaussianity in the cosmology with a domain wall era is thus enhanced by a factor of $\mathcal{T}^{-2}$ relative to the standard cosmology (with the same $f_{\text{NL}}$). However, the local bispectrum is still given by
\begin{equation}
\langle\zeta_{\vec{k}_1}\zeta_{\vec{k}_2}\zeta_{\vec{k}_3}\rangle = (2\pi)^3\delta(\vec{k}_1 + \vec{k}_2 + \vec{k}_3)\times \frac{6}{5}f_{\text{NL}}^{loc} \frac{\mathcal{P}^2_{\zeta}}{(k_1 k_2 k_3)^3}\left(\frac{k_1^2}{k_2 k_3} + \frac{k_2^2}{k_1 k_3} + \frac{k_3^2}{k_1 k_2}\right)
\end{equation}
if one assumes the naive parameterization $\zeta(x) = \zeta_g(x) + \frac{3}{5}f_{\text{NL}}^{loc}(\zeta^2_g(x) - \langle\zeta^2_g(x)\rangle)$ to be the form of the local correction to the Gaussian perturbation $\zeta_g$ \cite{Komatsu:2000vy}.
Since the walls don't affect the value of $f_{\text{NL}}^{loc}$ at large scales, the observable non-gaussianity today (when  $\mathcal{P}_\zeta\sim 10^{-9}$) is the same as in the standard cosmology.  The one potential enhancement could be due to the fact that the inflation scale can be closer to the Planck scale, and thus higher dimensional operators contributing to the potential and kinetic terms of the inflaton may be more significant (as in any model with a low enough cutoff). 
\chapter{Origin and Collapse of the Network}
\label{ch:Model}
\chaptermark{Origin and Collapse of the Network}

So far, we have treated the domain wall network as an inflating elastic solid to calculate late-time observables at large scales in a cosmology with a domain wall era, as the one depicted in Fig.(\ref{fig:CosmicHistory}). This macroscopic description is valid as long as the number of domains per Hubble volume is always large. As an existence proof, in this Chapter we give an example of a hybrid inflation model \cite{Linde:1993cn} that can realize the proposed cosmology. We show that there is a region of parameter space where a dense domain wall network produced at the end of inflation, quickly dominates the energy density for a few e-folding before collapsing and reheating the universe.

\section{Model}\label{sec:Model}
Consider the approximate $O(N)$ model introduced in Section \ref{ssec:Kubotani} 
\begin{equation}\label{eq:Potential1}
V(\chi_1,\cdots,\chi_N)=\frac{\lambda}{4}\left(\chi^{T}\chi-v^2\right)^2+ \frac{\kappa}{4}\sum\chi_i^4
\end{equation}
where we have organized the $N$ scalar fields into the vector $\chi\equiv (\chi_1,\cdots,\chi_N)^T$, and where $\lambda$ and $\kappa$ are coupling constants. For $\kappa\ll N\lambda$, the potential has an approximate global $O(N)$ symmetry which becomes exact as $\kappa\rightarrow0$. To produce domain walls at the end of inflation, we couple the $N$ ``waterfall fields'' in $\chi$ to the inflaton $\phi$ as follows
 \begin{equation}\label{eq:Potential2}
V(\chi_1,\cdots,\chi_N,\phi)=\frac{\lambda}{4}\left(\chi^{T}\chi-v^2\right)^2 + \frac{\kappa}{4}\sum\chi_i^4+\frac{g^2}{2}\phi^2\chi^{T}\chi
\end{equation}
Note, the inflaton has its own potential $V_\phi$, but we won't need to specify its form in what follows. The $O(N)$ symmetry in the above potential is spontaneously broken as $\phi$ rolls below the critical value $\phi_c\equiv m/g$ and the effective mass squared of $\chi$ becomes negative. This instability causes each $\chi_i$ field to roll down its potential and randomly settle at one of the available vacua in the model. The universe then divides into domains of initial size $L_0$, separated by domain walls of thickness $\delta_w$ and surface energy (or tension) $\sigma$.
\\
\indent To find the initial state of the network we need to understand the structure of the vacuum manifold. To this end, it is convenient to switch to the $N$-dimensional analog of spherical coordinates in field space. We write
$$\chi=\rho(\cos\theta_1, \sin\theta_1\cos\theta_2,\cdots,\sin\theta_1 \cdots \sin\theta_{N-2}\cos\theta_{N-1}, \sin\theta_1\cdots \sin\theta_{N-2}\sin\theta_{N-1})^T$$
where $\rho$ is a radial field and $\theta _1 , \theta _2 , \dots , \theta _{N-1}$ are $N-1$ angular fields, $\theta_{N-1}$ ranging from 0 to $2\pi$ and the rest ranging from 0 to $\pi$.  In these coordinates, the potential becomes
\begin{equation}\label{eq:PotentialSpherical}
V(\rho,\theta_1,\cdots,\theta_{N-1})= \frac{V_0N\lambda}{\kappa+N\lambda} + \frac{\lambda}{4}\rho^4 + \frac{\kappa}{4}\rho^4\Theta_{N-1}+ \frac{1}{2}\left(g^2\phi^2 - \lambda v^2 \right)\rho^2
\end{equation}
where $V_0\equiv\lambda v^4/4$. The constant term has been adjusted so that $V=0$ at its minimum, and for $0<\kappa\leq N\lambda$, it is always of order $V_0$. The angular dependence which explicitly breaks the $O(N)$ symmetry is encoded in the function $\Theta_{N-1}$, which can be obtained recursively using $\Theta_1(\theta)=\frac{1}{4}\left(3+ \cos4\theta\right)$ and
\begin{equation}
\Theta_{i}(\theta_1,\cdots, \theta_{i})=\cos^4\theta_{1} + \Theta_{i-1}(\theta_{2},\cdots,\theta_{i})\sin^4\theta_{1};\qquad\quad(\text{for } 2\leq i\leq N-1)
\end{equation}
The $\phi$-dependent, effective mass of $\rho$ (given by the last term in \ref{eq:PotentialSpherical}) is
\begin{equation}\label{eq:effectivemassrho}
m_{eff}^2(\phi)\equiv g^2\phi^2-\lambda v^2
\end{equation}
At early times when $\phi\gg \phi_c\equiv \sqrt{\lambda} v/g$, the effective mass is positive and the potential is minimized at the $O(N)$-symmetric point where $\rho=0$. Assuming $V_0\gg V_\phi$, there is a period of slow-roll inflation driven by the vacuum energy $V_0\simeq 3M_p^2H_i^2$ where $H_i$ denotes the nearly constant Hubble parameter during this inflationary period. Below the critical point, the effective mass becomes negative and the radial field can roll towards a new minimum at $\rho_{min}\equiv v \sqrt{N\lambda/(\kappa + N\lambda)}\sim \mathcal{O}(v)$ and $\phi=0$. Once the radial field reaches the bottom of the potential, inflation ends, the approximate $O(N)$ symmetry is broken spontaneously, and the low energy effective potential for the angular fields becomes
\begin{equation}\label{eq:PotentialSpherical2}
V_{low}(\theta_1,\cdots,\theta_{N-1})\sim \frac{\kappa}{4}v^4\left(\Theta_{N-1} -\frac{1}{N}\right)
\end{equation}
This potential is minimized (i.e. $V_{low}=0$) when each $\theta_i$ equals one of the following values:
\begin{eqnarray*}
\theta_{i}^{min}&=&\tan^{-1}(\sqrt{N-i}) \quad\text{or}\quad \pi- \tan^{-1}(\sqrt{N-i}),\quad\text{for}\quad i\neq N-1\\
\theta_{N-1}^{min}&=& \pi/4,\quad3\pi/4,\quad5\pi/4,\quad\text{or}\quad 7\pi/4
\end{eqnarray*}
so the vacuum manifold is a set of $n_v=4\times2^{N-1}=2^N$ discrete points corresponding to the corners of an $N$-cube.
\\
\indent This model allows for the formation of $N$ types of walls of different tension. For $N>2$, one can show that $N-1$ of these walls (the ones interpolating vacua across diagonals of the $N$-cube) have tensions that are more than twice the tension of the least energetic wall (the one interpolating adjacent vacua). So, the heavy walls always decay into the stable ones. Note, there are $N\times2^{N-1}$ stable walls, corresponding to the number of edges in the $N$-cube. Since this number is always even, a Hubble-sized domain wall network that forms in \textit{real} space, is roughly given by a (3D) cubic lattice of length $H^{-1}_i$ and lattice spacing $L_0$, where the initial domain size $L_0$ can be found by inspecting the dynamics near the critical point (we do this in section \ref{sec:DomainSize}). Moreover, the walls are joint together via sub-dominant defect structures: string-like junctions form when more than two distinct vacua cycle around a point in space, and monopole-like junctions form where these string-like junctions meet. As a result, an isotropic  stable structure can form inside the horizon, once the walls stretch under their own tension.
\\
\indent When the network forms, each of the $n_0$ domains per Hubble patch (i.e., each cell in the lattice) gets populated at random by one of the available vacua. One might then ask, what is the probability for an ``infinite'' domain to form once all of the domains with same vacua merge and the system stabilizes. To answer this, one can start at the center of a domain, where the probability of it being in one of the $2^N$ possible vacua, is $p=1/2^N$. Then, for this domain to double in size, one of the 26 surrounding domains must be in the same vacuum. Following this logic, the probability for a given domain to grow by a factor of $n$ ($n$ being an integer) is
\begin{equation}
P_n \leq 26\times25^{n-1}\times p^n = (26/25)(25p)^n
\end{equation}
Thus, in the model with $N=5$, for which $p=1/32$, there are no infinite domains (given by the limit where $n \rightarrow H_{i}^{-1}/L_0$), and the probability that any domain is more than a few times its original size $L_0$, is small (note, the upper bound of $P_n$ gets multiplied by a factor of $25/2^N\ll1$, whenever $n\rightarrow n+1$). We conclude that, as long as $N\geq5$, the initial percolation of the vacua can be ignored. We will show in section \ref{sec:DomainSize} that the number of vacua in the model is typically of order $10^{14}$; see Eq(\ref{eq:n0final}).
\\
\indent Before moving on to the next section, let us estimate the typical domain wall thickness and tension in this model. For $\kappa \ll N \lambda$ and $N\gg 1$, on can show that the vacuum energy under the potential barriers which generate the stable domain walls is of order $\kappa v^4/(4N^2)$, while the gradient energy is of order $(\nabla \chi_i)^2 \sim \delta_w^{-2} (v^2/N)$. Balancing these two energies then gives
\begin{equation}\label{eq:thickness}
\delta_w\sim \frac{2\sqrt{N}}{\sqrt{\kappa}v}
\end{equation}
and thus a wall energy per unit area of order
\begin{equation}\label{eq:sigma}
\sigma \sim \delta_w \frac{\kappa v^4}{4N^2}\sim \frac{\kappa^{1/2} v^3}{2N^{3/2}}
\end{equation}

\section{Initial Domain Size}\label{sec:DomainSize}
The initial number of domains per Hubble volume ($n_0$) must be large for the network to be able to dominate the energy density for a few e-foldings and then decay homogeneously, and also for the solid description of the network at large scales to be valid. To estimate $n_0$ in our model we must inspect the dynamics near the critical point, when the radial field becomes tachyonic. The tachyonic instability can be triggered as the inflaton $\phi$ classically rolls passed the critical point ($\phi_c$), or it can be induced by the interaction between quantum fluctuations in all fields. We now study these two regimes and show that a large $n_0$ is significantly more favorable in the quantum regime. This is expected since the dynamics near criticality in our model is typically dominated by quantum fluctuations, given that the primordial power can take values close to one.

\subsection*{Classical Regime}
In this regime, the classical rolling of $\phi$ dominates the dynamics at the transition. Setting $\phi(t)\simeq \phi_c - |\dot{\phi}|t$ in the effective mass squared of $\rho$, given in Eq.(\ref{eq:effectivemassrho}),
we see that (near the critical point)
\begin{equation}
|m_{eff}^2(t)|\simeq \omega^3 t
\end{equation}
where the scale $\omega\equiv (2g\sqrt{\lambda}v|\dot\phi|)^{1/3}$ parameterizes the rate of linear growth. Thus, quantum fluctuations in $\rho$, with $k^2<\omega^3t$, grow exponentially, while those with $k^2>\omega^3t$ keep oscillating in the quantum vacuum. Defining $z\equiv\omega t$ (with $z\gg k^2/\omega^2$), the average size of the growing fluctuations can be shown to scale as \cite{CorrelationLength, SpikeLyth2, SpikeLyth, Asaka:2001ez}
\begin{equation}\label{eq:Variance}
\langle \rho^2\rangle(z) \simeq \frac{\omega^2}{z}\exp\left(\frac{4}{3}z^{3/2}\right)
\end{equation}
The growth then stops when $\langle \rho^2 \rangle \sim v^2$, i.e, when $z= z_{end}\sim \left(\ln(v/\omega)\right)^{2/3}$. Using $P_{\zeta}(k_*)=H_i^4/(4\pi^2\dot{\phi}^2) \sim 10^{-9}/\mathcal{T}$  and $v \sim \sqrt{M_p H_i} \lambda^{-1/4}$, it is easy to show that
\begin{equation}\label{eq:mu}
\omega \sim 30 H_i \left(\frac{g^2 \sqrt{\lambda}\mathcal{T} M_p}{H_i}\right)^{1/6}\quad \longrightarrow \quad z_{end}\sim \left(\frac{1}{3}\ln\left(\frac{10^{-4}M_p}{g \lambda H_i \sqrt{\mathcal{T}}}\right)\right)^{2/3}
\end{equation}
so $z_{end}$ is typically of order a few, with a strong upper bound around $20$ \cite{SpikeLyth2}. Thus, the typical momentum of the enhanced modes, with $k^2\lesssim \omega^2 z_{end}$, will be of order $\omega$. This leads to the division of the universe into homogeneous domains of order\footnote{Note, if the transition was instantaneous, all modes with $k<\sqrt{\lambda}v$ would start to grow at $t=0$, and we would get $L_{cl}\sim 1/(\sqrt{\lambda}v)$.} $L_{cl} \sim 1/\omega$, separated by domain walls of thickness $\delta_w$, as given in Eq.(\ref{eq:thickness}). Imposing the thin-wall condition $\delta_w\ll L_{cl}$ for simplicity, we find that
the number of domains in a Hubble patch at the end of the phase transition is
\begin{equation}\label{eq:DWdensityC}
n_{0}\sim\left(\frac{H_i^{-1}}{L_{cl}}\right)^3 \sim 10^4\left(\frac{g^2\sqrt{\lambda}M_p \mathcal{T}}{H_i}\right)^{1/2}
\end{equation}
Thus, $n_0$ get suppressed when $\mathcal{T}\sim 10^{-9}$, that is, when quantum fluctuations are large and the primordial spectrum is order one. Note, decreasing the inflationary Hubble scale increases $n_0$. However, since $\dot{\phi}\propto H_i^2$, there is a point where the inflaton's speed becomes too low to outpace quantum effects (see Eq.(\ref{eq:Qdiff})) which, as we now discuss, change the dynamics near the critical point.

\subsection*{Quantum Regime}
In this regime, $\dot{\phi}$ is negligible when $\phi\simeq \phi_c$, and quantum fluctuations between all fields dominate the dynamics near the transition. The direction in field space where the growth of a linear combination of $\rho$ and $\phi$ is more likely to occur, can be found by expanding the potential about the critical point. Setting $\phi=\phi_c -r\cos\alpha$ and $\rho=r \sin\alpha$, the potential becomes
\begin{equation}\label{eq:ExpandedPotential}
V(r,\theta)= \frac{V_0N\lambda}{\kappa+N\lambda} - g \sqrt{\lambda}v r^3\cos\theta \sin^2\theta + \mathcal{O}(r^4) 
\end{equation}
Note that the dependence on the fields $\theta_1\cdots\theta_{N-1}$ appears only at sub-leading orders in $r$, and can thus be ignored. Minimizing Eq.(\ref{eq:ExpandedPotential}) with respect to $\theta$, the effective potential along the steepest descent is
\begin{equation}\label{eq:ExpandedPotential2}
V(r)= \frac{V_0N\lambda}{\kappa+N\lambda} - \frac{2}{3\sqrt{3}}g\sqrt{\lambda}v r^3 + \mathcal{O}(r^4)
\end{equation}
This leads to a negative curvature of order $V''(r)\sim -g\sqrt{\lambda}v r$, so the quantum fluctuations in $r$ with momentum $k$ are amplified if \cite{Dufaux:2008dn}
$k^2 \lesssim |V''|\sim g\sqrt{\lambda}v \sqrt{\langle r^2\rangle}\sim  g\sqrt{\lambda}v k$, where the last relation comes from the standard calculation of quantum fluctuations at short wavelengths (see, for example, \cite{MukhanovBook}). The typical momentum of the enhanced modes is thus of order $g \sqrt{\lambda}v$, which (as one might expect) is of the same order as the scale of the dimensionful coupling in the $r^3$ potential in Eq.(\ref{eq:ExpandedPotential2}). This leads to the division of the universe into homogeneous domains of order $L_{q} \sim 1/(g\sqrt{\lambda}v)$.
\\
\indent To be more precise, let us write $L_{q}= 1/(cg\sqrt{\lambda}v)$, for some constant $c$. We can then estimate $c$, using the assumption that the transition between classical and quantum regimes is smooth. The transition occurs when $\dot{\phi} = m^2_{eff}(t_{end})=\omega^2z_{end}$, that is, when the change in the inflaton's value, as it classically rolls near the critical point, equals the inverse Compton wavelength of $r$ at the end of its tachyonic growth. At this point, we must also have $L_{cl}=L_{q}$, which can be rewritten as $\dot{\phi}=\omega^2c/2$. Therefore, the matching requires that $c=2z_{end}$, where $z_{end}$ is of order a few; see Eq.(\ref{eq:mu}). Note, this is consistent with the analysis in \cite{Dufaux:2008dn}, where it was found empirically that $c\simeq10$ for $g^2=2\lambda = 2\times10^{-5}$ and $v= 10^{-3}M_p$ (for these values, our estimate give $c\simeq 9.5$).
\\
\indent Taking the thickness (as given in Eq.(\ref{eq:thickness})) of the walls to be small compared with the domain size -- $\delta_w L_q^{-1} \simeq 2 c g \sqrt{N\lambda/\kappa}\ll 1$ -- we find 
\begin{equation}\label{eq:n0final}
n_{0}\sim \left(\frac{H_i^{-1}}{L_q}\right)^3 
\sim 10^{14} 
\left(\frac{10^6 \text{GeV}}{H_i}\right)^{3/2}
\left(\frac{\delta_w L_q^{-1}}{0.1}\right)^3
\left(\frac{\kappa}{N\lambda}\right)^{3/2}
\left(\frac{\lambda}{0.1 }\right)^{3/4},
\end{equation}
which can be very large for a range of parameters. Lastly, we point out that quantum fluctuations dominate the dynamics when $L_q<L_{cl}$, that is, when 
\begin{equation}\label{eq:Qdiff}
H_i<\frac{c^{3}g^2\sqrt{\lambda}M_p}{10^4\sqrt{\mathcal{T}}} \quad\quad\quad(\text{Quantum beats Classical})
\end{equation}
which is a rather weak constraint on $H_i$.

\section{Network Evolution}\label{ssec:Evolution}
When domain walls form after inflation, they are some fraction of the energy density, and not necessarily dominant. In terms of the vacuum energy during inflation, $V_i \simeq 3 M_p^2 H_i^2$, the fraction of energy density in the walls is
\begin{equation}\label{eq:R}
\frac{\sigma L_0^{-1}}{V_i} \sim \frac{2 c g}{N} \sqrt{\frac{\kappa}{N\lambda}}
\end{equation}
where we have used Eq.(\ref{eq:sigma}), $L_0 = L_q 1/(c g \sqrt{\lambda} v)$ and $V_i \sim \lambda v^4 /4$.
If $N$ is large enough to form a non-trivial domain-wall network, it should eventually dominate.  The naive equation of state for domain walls is $w_{walls} = -(2/3) + v_{rms}^2$ \cite{KolbTurner, CosmicStrings}, but this doesn't take into account dissipation. As mentioned in \ref{ssec:NetworkEvolution}, the cosmic evolution of domain walls while they are sub-dominant can in principle be studied using simulations. The most recent state-of-the-art simulations \cite{Avelino:2008ve} suggest that domain walls, even when connected via junctions, enter a so-called scaling regime. In this regime, the walls disentangle, but the annihilation processes occur in such a way that the number of domains per Hubble volume remains \textit{fixed} as the universe expands ($L\propto H^{-1}\sim a^2$ during radiation domination). Note, the energy density in the network always dilutes at a slower rate than the energy density in the dominant component. If we write $L/L_0 \simeq H_i/H$, then 
\begin{equation}
\frac{\rho_{walls}}{\rho_{dominant}} 
\sim \frac{\sigma L^{-1}}{3 M_p^2 H^2}
\sim \frac{\sigma L_0^{-1}(H/H_i)}{3 M_p^2 H^2} 
\quad (\sim a^2 \text{ if radiation dominated})
\end{equation}
so the sub-dominant network will dominate the energy budget when this ratio is unity, or $H \sim \sigma L_0^{-1} / (3M_p^2 H_i)$. The number of e-folds elapsed during this intermediate period (assuming radiation domination) is then 
\begin{equation}
\N_{int} = \ln\left(\frac{a}{a_i}\right) = -\frac{1}{2}\ln\left(\frac{\sigma L_0^{-1}}{V_i}\right) 
\end{equation}
\indent Once the network dominates the energy density, the universe starts to inflate and we expect the residual wall velocities to decay rapidly due to the Hubble friction. Assuming that the network frustrates upon domination, we have $w_{walls}\simeq -2/3$ and $L\propto a$. Since $\rho_{walls}\sim a^{-1}$, it is easy to show that the number of e-folds during the domain wall era is 
\begin{equation}\label{eq:Nint}
\N_{dw} = 2\ln\left(\frac{\sigma L_0^{-1}}{V_i}\right) + \ln\left(\frac{V_i}{\rho_{rh}}\right)
\end{equation}
where $\rho_{rh}\sim T_{rh}^4$ is the energy density when the walls decay and reheat the universe to a temperature $T_{rh}$.  Note, if instead the network stays in the scaling regime during domain wall dominance, we expect the effective rigidity of the ``solid'' to be lower, and thus  the suppression of perturbations outside the horizon to be milder.

\section{Network Decay and Reheating}
The network must eventually collapse and reheat the universe. A simple way to make it collapse is by adding a term in the potential whose effect is to lower the energy of one of the $2^N$ degenerate vacua \cite{Zeldovich, Gelmini:1988sf, Chang:1998tb}. If the vacuum degeneracy gets broken by some amount $\Delta \mathcal{E}$, then the volume pressure from a domain in the true vacuum will start to affect the walls around it when the energy density $\rho$ in the universe gets diluted to values of order $\Delta \mathcal{E}$. More precisely, given a true vacuum domain of size $L^3$, there is a force $F\sim \Delta \mathcal{E} L^2$ acting on the surrounding walls of mass $M\sim\sigma L^2$ that separate it from the false vacuum regions. When $\rho\sim\Delta \mathcal{E}$, the domains populated by the true vacuum will start to grow at a rate given by $F/M\sim\Delta \mathcal{E}/\sigma\sim 1/L$. As the walls accelerate, they  collide with one another and become radiation\footnote{They also radiate particles as they accelerate \cite{CosmicStrings}.} \cite{Watkins&Turner}. As long as $L\ll H^{-1}$ during this reheating period, the network's collapse occurs in much less than a Hubble time, justifying the choice of an instantaneous transition when matching perturbations in Section \ref{sec:matching}. Assuming the field constituents of the walls are coupled to Standard Model particles, the resulting plasma of particles in each Hubble volume will decay and reheat the universe to a temperature $T_{rh}\sim (\Delta \mathcal{E})^{1/4}$.
\\
\indent During the domain wall era, the Hubble radius scales as $a^{1/2}$, while $L$ scales as $a$, so the number of walls per Hubble patch decreases as $a^{-3/2}$. Thus, the final number of domains left in a Hubble patch before reheating is
\begin{equation}\label{eq:nrh}
n_{rh} = n_0 e^{-\frac{3}{2}\N_{dw}} \sim n_0 \mathcal{T}^{3/5}
\end{equation}
where we used Eq.(\ref{eq:T}) to write $ e^{-\frac{3}{2}\N_{dw}}\sim \mathcal{T}^{3/5}$ (the oscillatory effects in Eq.(\ref{eq:T}) are always negligible compared to the exponential suppression). For the network to collapse everywhere in the universe, we must ensure that every one of these Hubble patches contains at least one true-vacuum domain. If $p$ is the probability that a single domain is in the true vacuum, then the probability that every Hubble patch in the universe contains at least one true-vacuum domain is
\begin{equation}\label{eq:Prh}
P_{rh} = (1 - (1-p)^{n_{rh}})^{N_p}
\end{equation}
where $N_p$ is the number of Hubble patches in the entire universe at reheating. Using $N_p \sim\left(\CH_{rh}/\CH_0\right)^3$, where $\CH_{rh}$ is the comoving Hubble parameter at reheating and $\CH_0$ is the comoving Hubble parameter today, we find 
\begin{equation}\label{eq:Np}
N_p \sim \left(\frac{T_0 T_{rh}}{H_0 M_p}\right)^3 \left(\frac{\pi^2 g_{rh}}{90}\right)^{3/2}
\sim 10^{65}\left(\frac{T_{rh}}{10^{10}\text{GeV}}\right)^3
\end{equation}
where $T_0$ is the temperature of the universe today, $H_{0}$ is the Hubble parameter today and $g_{rh}\sim 10^2$ is the number of relativistic degrees of freedom at reheating. For $N\gg1$ ($p=1/2^{N}\ll1$), one can then show that to get $P_{rh} \geq 1 -\epsilon$, with $\mathcal{\epsilon} \ll1$, one requires  
\begin{equation}\label{eq:nrhmin}
n_{rh} \geq n_{rh}^{*}\sim 150\times2^{N} + 3\times 2^N\ln\left(\frac{T_{rh}}	{10^{10}\text{GeV}}\right) - 2^N\ln\left(\frac{\mathcal{\epsilon}}{0.01}\right) \end{equation}
For $\delta_w L_0^{-1}\ll 1$, this bound becomes
\begin{equation}\label{eq:Hbound}
H_i < 10^{10}\text{GeV} \times \left(\frac{\delta_w L_0^{-1}}{0.1}\right)^2\left(\frac{150\times 2^5}{n_{rh}^*}\right)^{2/3} \left(\frac{\mathcal{T}}{10^{-7}}\right)^{2/5} \left(\frac{\lambda}{0.1}\right)^{1/2} \left(\frac{\kappa}{N\lambda}\right)
\end{equation}
where we used Eq.(\ref{eq:nrh}) with $n_0$ as in Eq.(\ref{eq:n0final}). Setting $N=5$, $\lambda = 0.1$, $\kappa=0.5$, $g =4\times 10^{-3}$, $H_i = 10^{6}$GeV and $T_{rh}=10^{10}$GeV, one gets a cosmology with $\mathcal{P}_{\zeta}(k_*)\simeq 0.02$ and an inflationary scale of order $10^{12}$GeV. Moreover, $\N_{int}\simeq 2$, $\N_{dw}\simeq 9$, $n_0\sim 10^{14}$ and $n_{rh}\sim 10^8$ in this point of parameter space.
\chapter{Conclusion}
\label{ch:Conclusion}
\chaptermark{Conclusion}
In this thesis, we have studied a cosmology with a short period of domain-wall domination after inflation. Using an effective description of the (frustrated) domain wall network as a solid, we found that the scalar and tensor power spectra are suppressed by a factor of $e^{-\frac{5}{2}\N_{dw}}$ relative to their initial magnitudes after inflation. Since the value of the scalar primordial power spectrum can be several orders of magnitude larger than the measured value today, bounds on the energy scale of inflation in single-field slow-roll inflation can be relaxed. As an existence proof, we gave an example of a model which produces a domain wall network with the needed features to realize the proposed cosmology. In this model, we can have $\mathcal{P}_\zeta = \mathcal{O}(10^{-2})$ with an inflationary energy scale as large as $10^{12}$GeV. 

\clearpage
\appendix
\chapter{Newtonian Perturbation of a Fluid}
\label{app:NewtonianPert}
\chaptermark{Newtonian Perturbation of a Fluid}

In this appendix, for completeness, we review the Newtonian theory of perturbations in a fluid-like universe. We consider the following cases: (1) a static universe with no gravity, (2) a static universe with gravity, (3) an expanding universe with gravity. We also review Jean's instability. 
\\
\indent Consider a non-relativistic fluid with mass density $\rho$, velocity $\ub$, pressure $P\ll\rho$, and let a fluid element have position $\rb$. We will work in cosmological time $t$. Conservation of mass implies that if there is a flux of momentum $\rho \ub$ through a surface enclosing a volume, then the total mass in the volume must change by the same amount and thus
\begin{equation}\label{eq:continuity}
\partial_t\rho = - \nabla\cdot(\rho \ub) \quad\quad\quad\text{(Continuity equation)}
\end{equation}
Now let us find ``$F=ma$'' for the fluid element. Note that we have chosen coordinates such that $\ub=\ub(\rb,t)$, $\rho=\rho(\rb,t)$ where $\rb$ is a fixed location in the ``lab frame'' and $\ub$ is the velocity of the fluid particle at position $\rb$ at the instant $t$. To compute the acceleration in these coordinates we use $\ub(\rb+d\rb, t + dt)=\ub(\rb,t) + dt(\partial_t\ub) + (d\rb\cdot\nabla) \ub$, so that $\mathbf{a}=d\ub/dt=\partial_t\ub + \ub\cdot\nabla\ub$. Including pressure gradients and gravitational forces on the fluid element we have
\begin{equation}\label{eq:Euler}
(\partial_t + \ub\cdot\nabla)\ub= -\frac{\nabla P}{\rho}- \nabla\Phi\quad\quad\quad\text{(Euler equation, ``F=ma'')}
\end{equation}
where $\Phi$ is the gravitational potential which is determined by

\begin{equation}\label{eq:Poisson}
\nabla^2\Phi=4\pi G \rho  \quad\quad\quad\text{(Poisson equation)}
\end{equation}

\paragraph*{Static universe with no gravity}
Let us perturb the fluid (we denote all background quantities with over-bars) in a static background without gravity ($\Phi=0$, $\bar{\rho}$ and $\bar{P}$ are constant and $\bar{\ub}=0$). The above equations then imply (to linear order)
\begin{equation}
\partial_t\delta\rho = - \nabla\cdot(\bar{\rho} \ub)
\end{equation}
\begin{equation}
\bar{\rho}\partial_t\ub= -\nabla \delta P
\end{equation}
$\partial_t$ of the first equation combined with the divergence of the second give
\begin{equation}
\partial_t^2\delta\rho- \nabla^2\delta P=0
\end{equation}
For adiabatic perturbations we have $\delta P= (dP/d\rho) \delta \rho= w \delta \rho$. We can then write the wave equation for perturbations
\begin{equation}
(\partial_t^2-c_s^2\nabla^2)\delta\rho=0
\end{equation}
where we have identified the equation of state $w$ with the speed of sound: $w=c_s^2$. We see that the perturbation \emph{fluctuates} with frequency $\omega=c_s k$.

\paragraph*{Static universe with gravity}
Now let us add gravity ($\Phi=\bar{\Phi} + \delta \Phi$). The effect on the equation of motion is to add a source term to the right hand side
\begin{equation}
(\partial_t^2-c_s^2\nabla^2)\delta\rho= \nabla^2\delta\Phi=4\pi G \bar{\rho}\delta\rho
\end{equation}
and we now have $\omega^2=c_s^2k^2 - 4\pi G \bar{\rho}$. For large $k$ we still have oscillations, but for $k$ smaller than the Jeans length ($c_s^2k^2<4\pi G\bar{\rho}\equiv c_s^2 k^2_J$), gravity takes over and the fluctuations \emph{grow exponentially} due to the imaginary frequency.

\paragraph*{Expanding universe with gravity}
Our physical coordinates $\rb$ are now related to \emph{comoving} coordinates $\xb$ by $\rb(t)=a(t)\xb$, and thus $\dot{\rb}=\ub=H \rb + \vb$. The first term is the Hubble flow and the second is the proper velocity $a\dot{\xb}$. We will also use spatial derivatives with respect to comoving coordinates so that $\nabla_\rb= a^{-1}\nabla_\xb$. Similarly, to make explicit the effects from expansion, we use time derivatives with fixed $\xb$ instead of $\rb$. These are related by
\begin{equation}
\left(\frac{\partial}{\partial t}\right)_\rb = \left(\frac{\partial}{\partial t}\right)_\xb - \left(\frac{\partial}{\partial t}\right)_\rb\nabla_\xb=\left(\frac{\partial}{\partial t}\right)_\xb - H \xb\nabla_\xb
\end{equation}
We can then substitute $\nabla\rightarrow \nabla/a$, $\partial_t\rightarrow \partial_t - H \xb\nabla$, $\rho\rightarrow\bar{\rho}(1+\delta)$ and $u\rightarrow H a\xb + \vb$, where $\delta\equiv \delta\rho/\bar{\rho}$ is the density contrast, into Eq.(\ref{eq:continuity}), Eq.(\ref{eq:Euler}) and Eq.(\ref{eq:Poisson}) to obtain the continuity, Euler and Poisson equations with expansion included, respectively. After doing this, the continuity equation becomes
\begin{equation}\label{eq:continuityExpanding}
\left[\partial_t- H \xb\nabla\right]\left[\bar{\rho}(1 + \delta)\right] + \frac{1}{a}\nabla\cdot\left[\bar{\rho}(1 + \delta)(H a\xb + \vb)\right] =0
\end{equation}
It is easy to show that this reduces to 
$$\frac{\partial \bar{\rho}}{ \partial t} +  3 H  \bar{\rho} = 0$$
at zeroth order and to 
$$\dot{\delta} =  - \frac{1}{a} \nabla \cdot \vb$$
at first order (here the overdot denotes derivatives with respect to time $t$). Similarly, the Euler equation becomes
\begin{equation}\label{eq:EulerExpanding}
(\partial_t + \ub\cdot\nabla)\ub= -\frac{\nabla P}{\rho}- \nabla\Phi
\end{equation}
from which we see that for $\delta P = \delta \Phi = 0$ the proper velocity redshifts as $\vb \sim 1/a$. Lastly, the Poisson equation simply becomes 
\begin{equation}\label{eq:PoissonExpanding}
\nabla^2\delta \Phi=4\pi G \bar{\rho} \delta  \quad\quad\quad\text{(Poisson equation)}
\end{equation}

\paragraph*{Jeans' Instability}
One can combine the time derivative of Eq.(\ref{eq:continuityExpanding}), the divergence of Eq.(\ref{eq:EulerExpanding}) and Eq.(\ref{eq:PoissonExpanding}) to obtain
\begin{equation}\label{eq:JeansInstability}
\ddot{\delta} + 2 H \dot{\delta} - \frac{c_s^2}{a^2} \nabla^2 \delta = 4 \pi G \bar{\rho} \delta
\end{equation}
which is a single second order differential equation for the density fluctuation $\delta$. Note, the Jeans' scale is still $k_J = \sqrt{4 \pi G}/c_s$ but now $\bar{\rho}$ and $c_s$ are time-dependent due to the universe's expansion. Note also the presence of the Hubble friction term $2H \dot{\delta}$ which implies that: i) For lengths smaller than the Jeans' length, fluctuations oscillate with decreasing amplitude and ii) For scales larger than the Jeans' length, fluctuations undergo a power-law growth rather than the exponential growth experienced in the static case.

\chapter{Effective Field Theory of Broken Spatial Diffeomorphisms}
\label{app:EFT}
\chaptermark{Effective Field Theory of Broken Spatial Diffeomorphisms}
Fluctuations in the relativistic elastic solid (see Chapter \ref{ch:EvoPert}), can also be understood in the following effective field theory (EFT) approach, where fluctuations in the inflating solid are described by fluctuations about an FRW background in an EFT with spontaneously broken spatial diffeomorphisms (see \cite{Junpu} for a model of ``solid inflation'' where the background instead approaches a quasi de Sitter space with $w\simeq-1$, and where anisotropic stress gradients are negligible). This EFT has three Goldstone fields, one for each broken spatial diffeomorphism. Perturbations about their vacuum expectation values correspond to phonon fields. Decomposing the phonon 3-vector into one scalar and two vector components, and choosing a gauge where all perturbations are in the metric, the Goldstone modes get eaten by the graviton. The graviton then becomes a massive spin-2 particle with 5 healthy polarizations: the 1 scalar and 2 vector modes from the Goldstone sector plus the 2 tensor modes. Note, because the Goldstones are essentially ``ruler fields'' which select a preferred frame, perturbations are expected to relax towards the background state even at superhorizon scales. In contrast, in the effective field theory of inflation, only time diffeomorphisms are spontaneously broken and the Goldstone field is a ``clock'' field (in this case there are only 2 tensor and 1 scalar propagating degrees of freedom \cite{Cheung:2007st}).
\\
\indent After writing down all possible operators compatible with underlying symmetries, the most generic action for fluctuations around an FRW background with spatial diffeomorphisms breaking is \cite{Lin:2015cqa}
\begin{equation}\label{eq:EFT}
S=\int d^4x\sqrt{-g}\left[\frac{1}{2}M_p^2 R- 3 M_p^2\left(H^2+\dot{H}\right)+M_p^2a^2\dot{H}g^{ii} - M_p^2 M_2^2\bar{\delta} g^{ij}\bar{\delta}g^{ij}+\cdots\right]
\end{equation}
where $\bar{\delta} g ^{ij}$ is the traceless part of the spatial component of the perturbed FRW metric, written as $g^{ij}=a^{-2}\delta^{ij} + \frac{1}{3}\delta g^{kk}\delta^{ij}+ \bar{\delta} g ^{ij}$. $M_2$ is a time-dependent function encoding differences between possible UV completions. Note, there is only one unique quadratic operator and the higher order ones can be dropped since they redshift faster. After performing a full perturbation analysis and integrating out the 3 scalar and 2 vector non-dynamical metric degrees of freedom, the quadratic action of all 5 polarizations can be found. The dispersion relations for each of these dynamical modes can then be read directly from the action. In general, these relations are complicated functions of $k$ (see \cite{Lin:2015cqa}), however, they take simple relativistic forms in the small wavelength limit of the excitations. In this limits, the scalar and vector speeds of sounds can be found to be
\begin{equation}\label{eq:speeds}
\begin{split}
c_s^2 &\simeq \frac{1}{3} + \frac{2 \epsilon}{3} + \frac{8 M_2^2}{3 a^4 H^2 \epsilon}, \quad\quad\quad\quad (k^2\gg a^2 H^2\epsilon)\\
c_v^2 &\simeq 1 + \frac{2 M_2^2}{a^4 H^2 \epsilon}, \quad\quad\quad\quad\quad\quad\quad\quad (k^2\gg a^2 H^2\epsilon)\\
\end{split}
\end{equation}
where again, we have assumed $\epsilon$ to be constant. 
From the above equations, it then follows that  
\begin{equation}\label{eq:speedsrelation}
\begin{split}
c_s^2 - \frac{4}{3}c_v^2 &= w, \quad\quad\quad\quad (k^2\gg a^2 H^2 \epsilon)
\end{split}
\end{equation}
These expressions obtained using the EFT approach are the same as those obtained in eq.(\ref{eq:SoSRel}) for the relativistic elastic solid. 


\bibliographystyle{IEEEtran}
\bibliography{IEEEabrv,thesis} 


\end{document}